\documentstyle[prd,aps,epsf]{revtex}

\input{epsf}

\begin{document}

\sloppy
\setcounter{page}{0}
\setcounter{figure}{0}
\thispagestyle{empty}

\begin{flushright}
hep-ph/9706425
\end{flushright}
\vspace{3.5cm}

\begin{center}
{\Large
{\bf  A  QCD  SPACE-TIME ANALYSIS OF
\\ ${}$
\\ ${}$
QUARKONIUM FORMATION AND EVOLUTION
\\ ${}$
\\ ${}$
IN HADRONIC COLLISIONS}
}
\end{center}
\bigskip

\begin{center}
{\large {\bf Klaus Geiger}}
\vskip 0.2in
{\large {\em Physics Department, Brookhaven National Laboratory,
Upton, NY 11973, USA }}
\medskip

e-mail: klaus@bnl.gov
\end{center}
\bigskip
\bigskip
\bigskip

\begin{abstract}
The production of heavy quarkonium as $Q\bar{Q}$ bound-states in
hadron-hadron collisions is considered within the framework of a
space-time description, combining parton-cascade evolution with a 
coalescence model for bound-state formation. The `hard' production 
of the initial $Q\bar{Q}$, directly or via gluon fragmentation and 
including both color-singlet and color-octet contributions, is calculated 
from the PQCD cross-sections.  The subsequent development of the 
$Q\bar{Q}$ system is described within a space-time generalization of
 the DGLAP parton-evolution formalism in position- and momentum-space.
The actual formation of the bound-states is accomplished through 
overlap of the $Q\bar{Q}$ pair and a spectrum of quarkonium 
wave-functions. This coalescence can only occur after sufficent 
gluon radiation reduces the $Q\bar Q$ relative velocity to a value 
commensurate with the non-relativistic kinematics of these bound 
systems. The presence of gluon participants in the cascade then is 
both necessary and leads to the natural inclusion of both 
color-singlet and color-octet mechanisms.  The application of this 
approach to $pp$ ($p\bar{p}$) collisions from $\sqrt{s}=$ 
30 GeV - 14 TeV reveals very decent agreement with available data 
from ISR and Tevatron - without the necessity of introducing fit parameters.
Moreover, production probabilities are calculated for a complete 
spectrum of charmonium and bottonium states, with the relative 
significance compared to open charm (bottom) production.  An analysis 
of the space-time development is carried through which sheds light on 
the relevance of gluon radiation and color-structure, suggesting a 
correponding experimental investigation.
\end{abstract}
\vskip 1.5in

\leftline{
}

\newpage

\section{INTRODUCTION}
\label{sec:section1}
\bigskip

The experimental analysis of quarkonia production in high-energy particle
collisions has brought in the last couple of years exciting discoveries,
both in $p+\bar{p}$ collisions at the Fermilab Tevatron \cite{tevatron}, and
in $Pb+Pb$ collisions at the CERN SPS \cite{Gonin}. In the first case, the
 interpretation of the Tevatron data \cite{Mangano93} of quarkonium production
in $p+\bar{p}$ at $\sqrt{s}=1800$ GeV, revealed a drastic disagreement with
expectations from theoretical calculations that described lower-energy data
very well.  Not only was the quarkonium  yield underestimated by a
far more than an order of magnitude (in particular for $J/\psi$ and $\psi'$), 
but also the observed
$p_\perp$-dependence of the cross-sections was very different from what was
predicted earlier. This has become known as the `CDF anomaly'.
In the second case, a significant suppression of
$J/\psi$ production observed in the CERN heavy-ion data \cite{Gonin} of
$Pb+Pb$ at $E_{beam}=158$ A GeV ($\sqrt{s} = 17$ A GeV), implied that a naive
extrapolation from experiments with lighter nuclei substantially
overestimated the $J/\psi$ production.
Clearly, these two discoveries concerning quarkonium
production, namely, `enhancement' in $p+\bar{p}$, and
`suppression' in $Pb+Pb$ have rather different origin.  However, if one is to
comprehend the underlying mechanisms in ion-ion applications from anything
close to fundamental QCD, then both aspects have to be considered.  In the
case of $p+\bar{p}$ (and also $e^+e^-$) vital theoretical progress based on
QCD's first principles has been made \cite{RevBraaten,RevMangano} in the last few
years. In the case of heavy-ion collisions
the degree of theoretical insight is still perhaps more suspect, at present mainly
covering ad-hoc prescriptions or somewhat oversimplified models,
however with promising exceptions \cite{kharzeev}.
\medskip

To set the stage, I recall that
the recent achievements in understanding quarkonium production 
in hadronic collisions at high
energy and momentum transfer were essentially stimulated by two
`observations' inferred from Tevatron data:
First, gluon fragmentation (and not direct $Q\bar{Q}$ production)
appears to be the dominant source of prompt quarkonia at high energies,
i.e., a process that involves the high-$p_\perp$ production of a virtual gluon,
followed by its subsequent evolution and decay into $Q\bar{Q}$.
Second,
a substantial fraction of quarkonium production at very high energy
arises from intermediate color-octet fluctuations during the evolution into a
final-state color-singlet $Q\bar{Q}$ pair, a contribution which 
is small at ISR energies but becomes very prominent at Tevatron energy
and beyond.
\medskip

Previous to these observations, calculations of charmonium production were  
limited to the use of the so-called color singlet model \cite{CSM1,CSM2,CSM3}
(for a recent review of heavy quarkonium production, see  \cite{Schuler}).
A theoretical breakthrough was achieved by Bodwin, Braaten, and Lepage 
\cite{BBL}, who developed a rigorous QCD formalism
by marrying Perturbative QCD (PQCD) and an effective field theory
within  Non-relativistic QCD (NRQCD). 
These authors combined in  systematic manner the short-distance, relativistic physics
of $Q\bar{Q}$ production and evolution, with the non-perturbative, supposedly
non-relativistic dynamics of the formation of hadronic $Q\bar{Q}$
bound-states.  This novel theory \cite{Beneke} exploited
some unique features of the quarkonium bound-states.  Firstly, the
high-$p_\perp$ production of a $Q\bar{Q}$ pair, directly  by a hard parton collision
or indirectly via gluon fragmentation, is a truly perturbative
process with the heavy quark mass $M_Q$ naturally bounding the typical momentum
scale from below.  Secondly, the internal $Q\bar{Q}$ dynamics is inherently
non-relativistic, since in order to have a finite probability for producing a
$Q\bar{Q}$ bound-state, the pair necessarily must have very small relative
momentum, despite potentially large total momentum.
\medskip

Quantitatively, an important ingredient is the presence in quarkonium physics
of several energy (or momentum) scales, well separated if the relative
velocity ${\bf v}$ of the $Q\bar{Q}$ satisfies $|{\bf v}| \equiv v \ll
1$. The three important scales \cite{BBL,Cho1} are
(i) the heavy mass $M_Q$, setting the lower
scale for the relativistic $Q\bar{Q}$ production with typical momenta 
$ \, \lower3pt\hbox{$\buildrel >\over\sim$}\, M_Q$ and small-size configuratons
$\sim 1/M_Q$, (ii) the momentum $M_Q v$, specifying the scale for bound-state
formation with typical momenta $\sim M_Q v$ and mean bound-state size $\sim
1/M_Qv$, and (iii) the kinetic energy $M_Qv^2$, which if small ensures that a
nonrelativistic treatment of the bound-state may be reasonable
\cite{RevQuigg,Eichten78}.  If $v \ll 1$ then $M_Q \gg M_Q v \gg M_Qv^2$ and a double
series expansion in the strong coupling $\alpha_s$ and the velocity
parameter $v$ is meaningful.  Since $v\propto 1/\ln M_Q$, typical values
for the heavy quark velocities are $v_{c\bar{c}} = 0.23$ and 
$v_{b\bar{b}} = 0.08$.
Bodwin, Braaten,and Lepage \cite{BBL} demonstrated that one may 
then really use  PQCD to compute the various color-singlet and 
color-octet contributions of different
$Q\bar{Q}$ Fock state components to quarkonium production. The latter,
however, are converted into formation probabilities for quarkonium only after
multiplication by  unknown non-perturbative
matrix-elements for the $Q\bar{Q}$ conversion into physical states.
Nonetheless, it triggered several works addressing te application to
$e^+e^-$ physics \cite{Cho2} and $p\bar{p}$ collisions \cite{Cho1,Cacciari}. 
In the context of aforementionened  `CDF anomaly', it was shown \cite{Cho1} that 
by estimating the color-singlet contributions from quarkonium potential models,
and by fitting the color-octet contributions  to experimental
data,   the addition of the  {\it color-octet mechanism} can impressively resolve 
the disagreement of the $p+\bar{p}$ Tevatron data with previous 
predictions within the {\it color-singlet model} only.
\footnote{
It should be noted that the so-called {\it color evaporation model} with an early 
history \cite{Fritzsch}, was also capable of including octets but is too
severely constrained by its assumptions \cite{RevBraaten} to have a
useful applicability.
}
\smallskip  
\begin{figure}
\vspace{-3.0cm}
\epsfxsize=490pt
\centerline{ \epsfbox{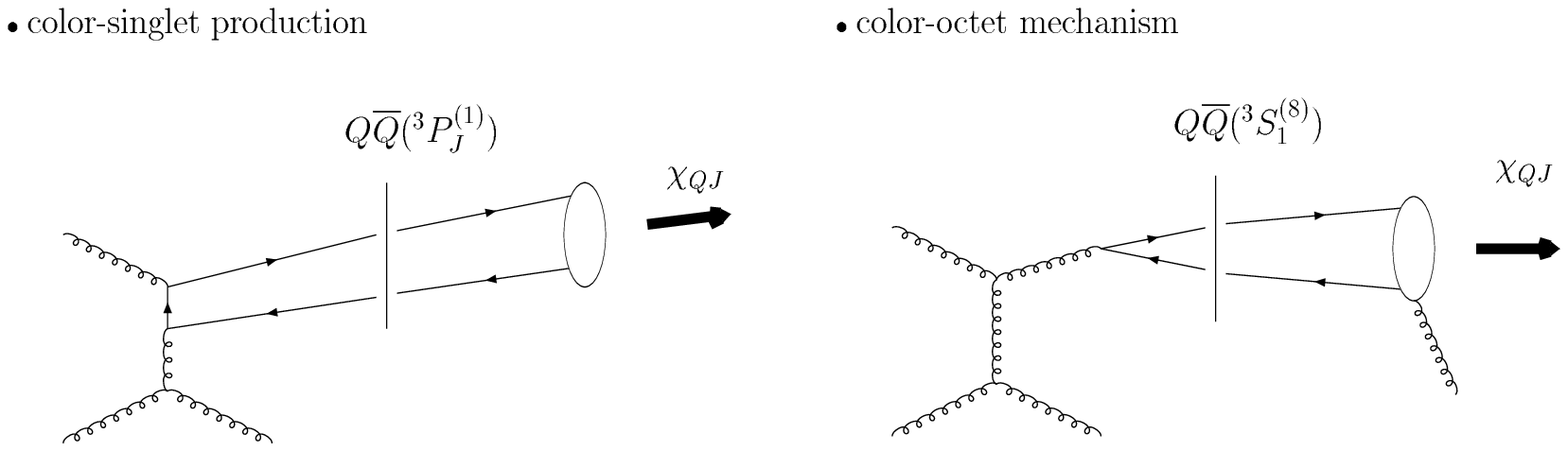} }
\vspace{-14.0cm}
\caption{
         Feynman graphs that contribute to $\chi_{Q\,J}$ production at
         $O(\alpha_s^3)$ through color-singlet and color-octet mechanisms.
         The initial $Q\overline{Q}$ is produced in a $^3P_J^{(1)}$ state, 
         respectively
         a $^3S_1^{(8)}$ state, at short distances and hadronizes into a 
         final-state $\chi_{Q\,J}$ bound-state at large distances.
         \label{fig:fig1}
         }
\end{figure}
\bigskip

Nonetheless, some critical remarks are at hand here.
The dictates of experimental data aside, the physical justification of the
color-octet mechanism  remains unsatisfactory.  To clarify this
statement and to motivate the aims of this paper, it is useful to draw a
somwhat idealistic picture, to be improved on later in this work, of
color-singlet and color-octet $Q\bar{Q}$ production in hadron-hadron
collisions.  
In Fig. 1a a typical {\it color-singlet} Feynman graph that
yields a $\chi_{Q\,J}$ is depicted. 
The $O(\alpha_s^3)$ hard scattering of the incoming
partons produces at a short-distance scale an outgoing color-less $Q\bar{Q}$
pair in a $^3P_J^{(1)}$ state plus a gluon.  The heavy $Q$ and $\bar{Q}$ fly
out from the initial collision point, almost on mass-shell, eventually
forming, at large distances, a hadronic $\chi_{Q\,J}$ bound-state.  
In Fig. 1b, a  {\it color-octet} $Q\overline{Q}$ pair in a  $^3S_1^{(8)}$ state
is created by the hard parton scattering, and the pair is assumed to color-neutralize
far away from the collision point by emitting an extremely  long-wavelength gluon
which takes away the color, but virtually no energy or momentum.
The heavy quark pair then may transform again into a  $\chi_{Q\,J}$ state.  

Whereas the color-singlet production is clearly plausible, the color-octet
mechanism seems to contradict basic quantum mechanics. The assumption that an
ultra-soft gluon takes away the color, but virtually no energy or momentum is
just wrong! Such a gluon cannot carry away any quantum numbers within a
finite time.  The emitted gluon must be sufficiently hard if it is to both
neutralize the color-octet $Q\bar{Q}$ and separate from the $Q\bar{Q}$ as an
individual and incoherent quantum before the $Q\bar{Q}$ pair hadronizes.  
Since the
typical formation time of a $Q\bar{Q}$ bound-state is $t_{form} \sim 1/M_Qv
\, \lower3pt\hbox{$\buildrel >\over\sim$}\, 1$ $fm$, the gluon must have at
least an energy  $\omega_g = O(1\;GeV)$ if it is emitted at early time, and even
more if it is radiated later. Such a gluon is certainly not a long-distance
fluctuation with negligible energy.  Hence, although the intervention of the
color-octet mechanism resolves the Tevatron data paradox, I am inclined to
pose, provocatively, the question: 
Can the ad-hoc motivation of the color-octet mechanism not be
extended towards a more physically intuitive
scheme, still consistent with the principles of PQCD and NRQCD?
In fact, the probability of directly producing sufficiently low relative 
velocity $Q\bar{Q}$ octet or singlet pairs for appreciable
coalescence to occur is very small. Thus, in a model which attempts to combine
both PQCD and NRQCD to generate quarkonium, moderately hard gluon radiation 
plays an integral part and both singlet and octet pairs enter straightforwardly.
The gluon radiation serves to reduce the difference momentum to a point 
where coalescence is more probable, and of course automatically enables both
singlet and octet mechanisms. 
\bigskip

This is the very point where the work of this paper begins. 
Inspired by the above discussion, I take steps
towards a  `microscopic' space-time QCD description of heavy
quarkonium production.
The `hard' production 
of the initial $Q\bar{Q}$, directly or via gluon fragmentation and 
including both color-singlet and -octet contributions, is calculated 
from the PQCD cross-sections.  The subsequent development of the 
$Q\bar{Q}$ system is described within a space-time generalization of
 the DGLAP parton-evolution formalism \cite{DGLAP1,DGLAP2} 
in position- and momentum-space.
The actual formation of the bound-states is accomplished through 
overlap of the $Q\bar{Q}$ pair and a spectrum of quarkonium 
wave-functions. 

The analysis in this paper is confined to  $pp$ ($p\bar{p}$)
collisions in the energy range $\sqrt{s}=$ 30 GeV - 14 TeV,
spanned by ISR, Tevatron and LHC beam energies.
Once  the
space-time structure of quarkonium production in `clean' $pp$ ($p\bar{p}$)
collisions has been better understood, one may step further to address 
the observed quarkonium suppression in `dirty' heavy-ion collisions,
where the space-time information of $Q\bar{Q}$ evolution in and interaction with
the nuclear medium is most essential.

In the approach that I advocate here and hope to present in a convincing way
in the remainder,
it is very natural to 
incorporate both color-octet and color-singlet mechanisms.
A produced $c\bar c$ pair can start in either 
color state, although of course they form charmonium only in the singlet.
As I shall show, gluon radiation plays so dominant 
a role in the final coalescence into a
bound-state that both singlet and octet states must be involved during the
$p\bar p$ or $pp$ evolution. 
At the same time it is of great importance
to estimate the unknown non-perturbative effects, embodied in the conversion
probabilities, from a dynamic coalescence picture \cite{coalescence} originally 
constructed for the production of deuterons from neutron and proton 
constituents of nuclear collisions. 

The very explicit
cascade space-time structure, for say a $p\bar p$ collision, permits  testing
for coalescence after repeated gluon emission from a heavy quark pair. 
The qualitative picture that underlies the present
phenomenological approach is inspired by the above-mentioned combination of
PQCD and NRQCD.  Accordingly, I also assume that the genealogy of a heavy
$Q\bar{Q}$ bound-state hadron involves three, in space-time well-separated
components: first, the production of an initial $Q\bar{Q}$ pair (directly or
via gluon fragmentation), second, its subsequent evolution into a final-state
$Q\bar{Q}$ pair, and third, the conversion of the $Q\bar{Q}$ pair into a
bound-state hadron $H_{Q\bar{Q}}$.  Imagine a heavy $Q\bar{Q}$ pair
emerging from a hard process of large momentum transfer scale, with relative
probability given by the appropriate hard QCD cross-sections.  This initial
off-shell $Q\bar{Q}$ pair, color-singlet or color-octet, then propagates in
space-time according to relativistic cascading, and simultanously, but
subject to the uncertainty principle, evolves in momentum space according to
the DGLAP evolution equations that describe the transmutation to the
final-state color-singlet $Q\bar{Q}$ in terms of real and virtual gluon
emission.  Finally, the probability for the $Q\bar{Q}$ pair to form a
hadronic bound-state $H_{Q\bar{Q}}$ is assumed to be given by the overlap of
the $Q$ qnd $\bar{Q}$ wave-functions with the bound-state wave-function.  
I will discuss and examine each of these three, essentially factorizable,
building blocks of the model in detail.

This novel approach to quarkonium production and bound-state
formation brings improvements on three distinct levels when compared to
previous work based on combining (lowest order) PQCD matrix-elements with
heavy quark effective theory in NRQCD:
\begin{description}
\item[1.]
A {\it dynamical cascade description} is employed that allows to trace the
detailed history of the evolution in both space-time and momentum space, from
the instant of initial $Q\bar{Q}$ production, through the propagation and
cascading of the pair and its gluonic offspring, and up to the formation of
the final bound-state hadron.

This is to be seen in strong contrast to the previous calculations, which
commonly were of a {\it static} nature, in the sense that no reference was made
to the microscopic time-development of the quarkonium system. Nor were
attempts made to account for `real' gluon emission or possible rescattering of
the quark-antiquark composite during its evolution.  In the present  approach on the
other hand, the evolution is governed by the dynamics itself, including
both `deterministic' gluon emission proceses and `statistical' scattering or
absorption processes. The space-time history is followed causally and
provides at any given time the trajectories, coordinates and interaction
vertices of the particles. 

\item[2.]
The {\it PQCD aspects} of the quarkonia were in earlier work described in terms
of lowest order perturbative matrix-elements for the production of a
quark-antiquark pair in a given state of angular momentum, spin and 
color-singlet  or color-octet \cite{Mangano93,RevBraaten}).  
It was assumed that the initially produced $Q\bar{Q}$
and the final bound-state forming $Q\bar{Q}$ were identical states
during the perturbative evolution, except for
the (in some works) included momentum scale dependence.  
Thus, there was no 
chance for the initial pair to change its spin or color state after
being produced by the emission of one or meore perturbative gluons.

In contrast, the present approach allows the initial $Q\bar{Q}$ 
to be produced statistically in
any spin and color combination and then to change its state by gluon
emission, absorption or scattering.  Thus, the final $Q\bar{Q}$ quantum state
can be altered and the eventual yield of bound-state hadrons is a statistical
outcome of the cascade development. Further, matrix-element calculations,
though in principle exact, are in practice limited to first and second order
in PQCD, suffering from the technical complications that enter with
increasing order in perturbation theory.  The cascade description, on the
other hand, is based on the DGLAP equations.  In principle the entire series
expansion in $\alpha_s$ is included by resummation, although admittedly in an
approximate fashion and limited to the leading log approximation.  Many
observables in high-energy $e^+e^-$ or $p\bar{p}$ collisions, for which even
second order matrix-elements calculations are insufficient, are reproduced
with high accuracy in the parton cascade description \cite{hamacher}.

\item[3.]
The probability for the {\it non-perturbative conversion} of the final state
$Q\bar{Q}$ state into a bound-state hadron $H_{Q\bar{Q}}$ with specific total
angular momentum $J$, orbital angular momentum $L$ and total spin $S$ has
been modeled throughout the literature by simply lumping all non-perturbative
dynamics into the value of the bound-state (radial) wave-function ${\cal
R}_0(r)$ for $L=0$ (or its derivative ${\cal R}_1'(r)$ for $L=1$) at zero
$Q\bar{Q}$ separation $r\equiv |{\bf r}_Q-{\bf r}_{\bar{Q}}| =0$.  The
underlying motivation for this approximation is borrowed from positronium
annihilation, assuming that the bound-state hadron emerges from the
annihilation of $Q$ and $\bar{Q}$, and that the probability for
$Q\bar{Q}\rightarrow H_{Q\bar{Q}}$ may be approximated by the wave-function
behavior uniquely at (or near) the origin.  Not only does this approach
neglect the finite width of the bound-state wave-function, but also implies
that the probability given by ${\cal R}_0$ or ${\cal R}_1'$ is independent of
time, hence independent of the actual spatial $Q\bar{Q}$ configuration along
time-dependent trajectories for the $Q$ and $\bar{Q}$.

In the present approach these shortcomings are absent, since it enables to evaluate
at any given time the probability for $Q\bar{Q}\rightarrow H_{Q\bar{Q}}$ as a
function of both relative sparations $r$ and relative momentum $k$ of the $Q$
and $\bar{Q}$, using the full expression for the wave-function overlap
integral (admittedly, model wave-function), rather than simply rely on the
value of the wave-function at zero $r$. This might be termed a dynamic
coalescence picture.  Moreover, I do not interpret the conversion process as
$Q\bar{Q}$ annihilation, but as a coalescence process, analogously to, e.g.,
deuteron production in nuclear physics. Simple but effective techniques have
been developed and tested in the literature \cite{coalescence}, which I adopt
and apply to the present problem. 
\end{description}
\medskip

The remainder of the paper is organized as follows.
Sec. II is devoted to introducing the theoretical framework 
corresponding to the qualitative picture drawn above.
Based on the factorization assumption of short-distance
PQCD production and evolution of heavy $Q\overline{Q}$
pairs, and the long-distance coalescence process with bound-state
formation, the total squared amplitude for quarkonium production
in hadronic collisions is obtained as the product of three functions
corresponding to $Q\overline{Q}$ production, parton cascade evolution, and
coalescence, repectively.
It is outlined how the resulting quarkonium cross-section can be calulated
by simulating the space-time development of the complete particle system
of a collision event  with Monte Carlo methods as implemented in the computer
program VNI \cite{ms44}.
In Sec. 3 the formalism is then specifically applied to
$pp$ ($p\bar{p}$) collisions in the energy range $\sqrt{s} = $ 30 GeV - 14 TeV,
corresponding to center-of-mass energies between ISR, Tevatron, and LHC experiments.
Comparison with experimental data from ISR and Tevatron for both charmonium and bottonium
production is made, and the underlying dynamics specific to the model calculations,
is examined in detail, in particular with respect
to the relevance of gluon radiation, the contributions of
color-singlet and color-octet $Q\overline{Q}$ systems, and the characteristics
of the space-time structure.
Sec. 4 is reserved for concluding remarks and a brief discussion
of future perspectives of the model, especially with respect to
application to heavy-ion collisions and the issue of quarkonium suppression
in dense nuclear matter.
\bigskip
\bigskip

\section{HEAVY QUARK  PRODUCTION, EVOLUTION \& COALESCENCE TO BOUND-STATES}
\label{sec:section2}
\bigskip

Focus of interest is the production of quarkonium states $H_{Q\overline{Q}}$
with definite quantum nubers $J,L,S$ on both the charm and bottom sector,
for which the following nomenclature is employed:
\begin{equation}
H_{c\overline{c}}(^{2S+1}L_J)\;\;=\;\;
\left\{
\begin{array}{r}
\; \eta_c \; (^1S_0)  \\ 
\; J/\psi\; (^3S_1)  \\ 
\; h_c \; (^1P_1)  \\ 
\; \chi_{c\,0}\; (^3P_0)  \\ 
\; \chi_{c \,1} \; (^3P_1)  \\ 
\; \chi_{c\,2} \; (^3P_2) \\ 
\end{array}
\right.
\;\;\;\;\;\;\;\;\;\;\;\;
H_{b\overline{b}}(^{2S+1}L_J)\;\;=\;\;
\left\{
\begin{array}{r}
\; \eta_b \; (^1S_0)  \\ 
\; \Upsilon\; (^3S_1) \\ 
\; h_b \; (^1P_1) \\ 
\; \chi_{b\,0}\; (^3P_0) \\ 
\; \chi_{b \,1} \; (^3P_1) \\ 
\; \chi_{b\,2} \; (^3P_2) \\ 
\end{array}
\right.
\;.
\label{def0}
\end{equation}
Following the outline of Sec. 1, the
total amplitude for the production of a quarkonium hadron $H_{Q\bar{Q}}$
in a  collision of beam particles $A,B$ may be written as a factorized ansatz. 
As schematically illustrated
in Fig. 2, the relevant amplitude  can then be represented as a product of
three components:
\begin{description}
\item[(i)]
the short-distance perturbative amplitude for production of a time-like
fluctuation, being either an off-shell heavy pair $c^\ast = Q\bar{Q}$ or
a virtual gluon $c^\ast= g$,  plus associated initial-state gluon- and 
$q\bar{q}$-production in a hard collision of partons in the incoming beam particles
$A,B$ (e.g. $p, \bar{p}$) with large momentum transfer $Q^2\gg\Lambda^2$;

\item[(ii)]
the perturbative evolution of the produced fluctuation $c^\ast$ into a
final-state color-singlet $[Q\bar{Q}]^{JLS\,(1)}$ pair with definite total
spin $S$, relative and total angular momentum $L$ and $J$, either by
$Q\bar{Q}\rightarrow [Q\bar{Q}]^{JLS\,(1)}$ ({\it direct production}), or
by $g\rightarrow [Q\bar{Q}]^{JLS\,(1)}$ ({\it gluon fragmentation}),
in both cases accompanied by emission of other gluons and $q\bar{q}$
pairs;

\item[(iii)]
the long-distance non-perturbative amplitude for the conversion of the 
final-state color-singlet $[Q\bar{Q}]^{JLS\,(1)}$ assumed to occur by 
coalescence, into a bound-state hadron $H_{Q\bar{Q}}(^{2S+1}L_J)$,
as categorized by (\ref{def0}).
\end{description}
\medskip

Hence, the formula for this factorized ansatz can be expressed as
\begin{eqnarray}
{\cal A}\left(AB\rightarrow H_{Q\bar{Q}} + X\right)
&=&
\;\;
{\cal A}\left(AB\rightarrow c^\ast + n \;g + n'\;q+ \overline{n}'\bar{q}\right)_{production}
\nonumber \\
& &
\nonumber \\
& &
\;\times\;
{\cal A}\left(c^\ast \rightarrow \left[Q\bar{Q}\right]^{JLS, \,(1)} + m\;g 
+ m'\;q+ \overline{m}'\bar{q} \right)_{evolution}
\nonumber \\
& &
\nonumber \\
& &
\;\times\;\;
{\cal A}\left( \left[Q\bar{Q}\right]^{JLS,(1)} \rightarrow 
H_{Q\bar{Q}}(^{2S+1}L_J)\right)_{coalescence}
\;,
\label{ansatz}
\end{eqnarray}
where $n g, n' q, \bar{n}' \bar{q}$ etc., are the additional gluons and
anti-quarks, respectively, being emitted from the incoming and outgoing
partons of the hard vertex, and during the cascade evolution (c.f. Fig.2). 
\bigskip

\begin{figure}
\epsfxsize=490pt
\centerline{ \epsfbox{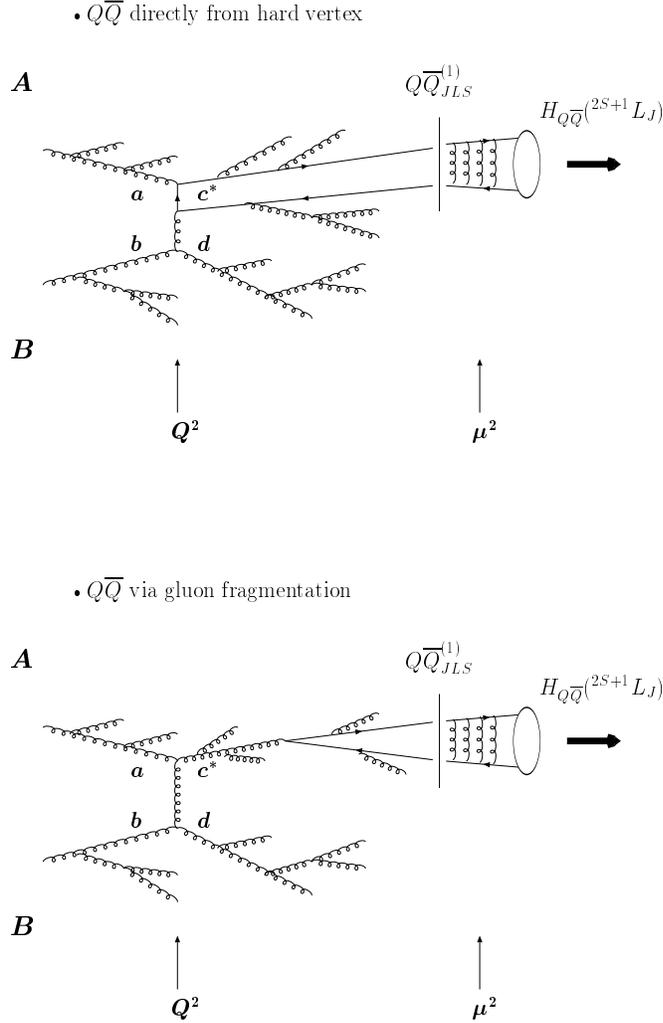} }
\vspace{-4.5cm}
\caption{
         Schematics of quarkonium production in the present approach:
         An initial $Q\overline{Q}$ pair is produced by a hard collision $ab\rightarrow c^\ast d$
         where $c^\ast=Q\overline{Q}$ (direct production) or $c^\ast=g$  with subsequent
         $g\rightarrow Q\overline{Q}$ (gluon fragmentation). The PQCD evolution
         includes real and virtual radiative processes off the incoming and outgoing
         partons. The  coalescence of the resulting final $Q\overline{Q}$ pair 
         yields a bound-state hadron $H_{Q\overline{Q}}$ with specific quantum numbers
         $J,L,S$.
         \label{fig:fig2}
         }
\end{figure}
\bigskip

In order to compute physical observables such as the 
total production cross-section, one needs to calculate the square of
the amplitude (\ref{ansatz}).
On account of the above factorization assumption, interference between
the three subamplitudes is ignored, and therefore
the total squared amplitude becomes a product of three squared
subamplitudes, each of which may be interpreted  probabilistically.
Henceforth, I denote these squared subamplitudes by 
\begin{itemize}
\item[(i)]
${\cal P}_{c^\ast}^2$ for the production process resulting in $c^\ast=Q\bar{Q}$,
or  $c^\ast=g$ with $g\rightarrow Q\bar{Q}$, 
\item[(ii)]
${\cal E}_{c^\ast}^2$ for the perturbative evolution in
space-time and momentum space, 
\item[(iii)]
${\cal C}_{JLS}^2$
for the coalecence mechanism and formation of a bound-state $H_{Q\bar{Q}}(^{2S+1}L_J)$.
\end{itemize}
Indicating
in their arguments the relevant momentum and space-time variables,
one can therefore express the square of (\ref{ansatz})
as follows:
\begin{eqnarray}
& &
\left| {\cal A}\left(AB\rightarrow H_{Q\bar{Q}} + X\right) \right|^2
\;\equiv\;
\left| {\cal A}\left(s,Q^2 \,|\, k_1, k_2, \overline{{\bf r}}_1,\overline{{\bf r}}_2, t
\right) \right|^2
\nonumber \\
& & \nonumber \\
& &
\;\;\;\;\;\;\;\;\;\;\;\;
\;=\;
\sum_{c^\ast= g, Q\bar{Q}}
\;
{\cal P}^2_{c^\ast}\left(\mu^2,Q^2,s \,|\, \overline{{\bf R}}_0, t_0 \right)
\;\times\;
{\cal E}^2_{c^\ast}\left(Q^2, \mu^2 
\,|\, k_1, k_2, \overline{{\bf r}}_1,\overline{{\bf r}}_2, t \right)
\;\times\;
{\cal C}_{JLS}^2
\left(\mu^2 \,|\, k_1, k_2, \overline{{\bf r}}_1,\overline{{\bf r}}_2, t \right)
\label{ansatz1}
\;.
\end{eqnarray}
The quantity $Q^2$ characterizes the hard scale
at the production vertex, $\mu^2=O$(1-2 GeV) is the mass scale that marks the boundary
between perturbative and non-perturbative regimes, and $\sqrt{s}$ is the
center-of-mass energy of the colliding beam particles $A$, $B$.
The four-momenta and mean trajectories of the $Q$ and $\overline{Q}$, 
respectively, are defined by 
\begin{equation}
k_i^\mu\;=\; (k^0_i, {\bf k}_i)
\;\;\;\;\;\;\;\;\;
r_i^\mu\;=\; (r^0_i, {\bf r}_i)
\;\;\;\;\;\;\;\;\;
{\bf \overline{r}}_i
\;=\;
\langle \;{\bf r}_i(t) \;\rangle
\;,
\end{equation}
where $i=1,2$ for $Q$ and $\bar{Q}$, respectively, and,
\begin{equation}
t\;=\;
\frac{1}{2} \left( r_1^0+r_2^0\right)
\;\;\;\;\;\;\;\;\;
{\bf \overline{R}}_0
\;=\;
\frac{1}{2} \left( {\bf \overline{r}}_1+{\bf \overline{r}}_2\right)_{t=t_0}
\end{equation}
is the time $t$, referring to the specific Lorentz frame in which one wishes to
describe the dynamical evolution, while $t_0$ and ${\bf \overline{R}}_0$ labels the 
point of time and mean  position at which the hard collision occurs.
\smallskip

Separate discussions of the three functions ${\cal P}^2$, ${\cal E}^2$ and
${\cal C}^2$ are assigned to the next three subsections.  The total
production cross-section for quarkonium mesons of type $H_{Q\bar{Q}}$ is then
obtained, in Section IID, by summing over all contributing channels and
multiplying ({\ref{ansatz1}) with the proper flux- and phase-space factors.
\bigskip

\subsection{Production}
\medskip

The  square of the production amplitudes 
$
{\cal A}\left(AB\rightarrow g^\ast + n \;g + n'\;q+ \overline{n}'\bar{q}\right)_{production}
$
in (\ref{ansatz1})
gives the probability for materializing a time-like fluctuation $c^\ast =Q\overline{Q}$
or $c^\ast=g$ with $k^2 \approx Q^2$ in a hard collision of partons
$a,b = g,q,\bar{q}$ at the hard scale $Q^2$ set by the momentum transfer at
the hard vertex.  As illustrated in Fig. 3, the squared amplitude can
be calculated from the usual $2\rightarrow 2$ hard-scattering amplitudes for
$ab \rightarrow c^\ast d$, multiplied by the space-like Sudakov formfactors
$S_{a,b}$ for the incoming partons $a$ and $b$ and the time-like formfactor
$T_d$ for the outgoing parton $d$, resumming the contributions of real and
virtual gluon radiation. The initial-state partons are weighted with the parton structure
functions $f_{a/A}$ and $f_{b/B}$ of the beam particles $A,B$. Hence, the
production probability is
\begin{eqnarray}
& &
{\cal P}^2_{c^\ast}\left(\mu^2,Q^2,s \,|\, \overline{{\bf R}}_0, t_0 \right)
\;\equiv\;
\left|
{\cal A}\left(AB\rightarrow c^\ast + n \;g + n'\;q+ \overline{n}'\bar{q}\right)_{production}
\right|^2
\nonumber \\
& &
\nonumber \\
& &
\;\;\;\;\;\;\;\;\;\;\;\;\;\;\;
\;=\;
\sum_{ab} \int dx_a\int dx_b  \,f_{a/A}(x_a,\mu^2) \; S_a(x_a,\mu^2,Q^2\,|\, t_0)
\;\;
\,f_{b/B}(x_b,\mu^2) \; S_b(x_b,\mu^2,Q^2 \,|\, t_0)
\nonumber \\
& &
\nonumber \\
& &
\;\;\;\;\;\;\;\;\;\;\;\;\;\;\;
\;\;\;\;\;\;\;\;\;\;\;\;\;\;\;
\;\;\;\;\;\;\;\;\;\;\;\;\;\;\;
\;\times\;
\left| A(ab\rightarrow c^\ast d)\right|^2(\hat{s},Q^2\,|\,\overline{{\bf R}}_0, t_0)
\;\times\;T_d(Q^2,\mu^2\,|\,t_0)
\;.
\label{A1}
\end{eqnarray}
where the sum is over all contributing parton channels, and 
the hard collision $ab\rightarrow c^\ast d$ is assumed to occur at
time $t_0$ and mean position $\overline{{\bf R}}_0 =\langle\; {\bf \vec{R}}_0
\;\rangle$. Furthermore,
\begin{equation}
s\;=\;(P_A+P_B)^2 
\;\;\;\;\;\;\;\;\;\;
\mu^2 \, \lower3pt\hbox{$\buildrel >\over\sim$}\,1\;GeV^2
\;\;\;\;\;\;\;\;\;\;
Q^2 \;\equiv\; p_\perp^2 
\label{def1}
\;.
\end{equation}
are, in that order, the total center-of-mass energy squared, the
factorization scale that marks the perturbative boundary below which the
non-perturbative dynamics is absorbed in the structure-functions, and the
momentum transfer scale at the hard vertex of the colliding partons $a$, $b$,
set by the relative transverse momentum of the outgoing parton jets.  The
Mandelstam variables at the parton level, $\hat{s},\hat{t}, \hat{u}$, and the
lightcone fractions $x_{a,b}$ of the colliding partons $a$, $b$ are defined
as usual
\begin{equation}
\hat{s}\;=\;x_ax_b\;s
\;\;\;\;\;\;
\hat{t}\;=\;\;-x_b\,p_\perp \,\sqrt{s}\, e^{y_2}
\;\;\;\;\;\;
\hat{u}\;=\;\;-x_a\,p_\perp \,\sqrt{s}\, e^{y_2}
\;\;\;\;\;\;\;\;\;\;
x_{a,b}\;=\;\frac{(k_0+k_z)_{a,b}}{\sqrt{s}}
\label{def2}
\;,
\end{equation}
where the incoming parton momenta  are parametrized as 
\begin{equation}
k^\mu_{a}\;=\; x_a\frac{\sqrt{s}}{2}\,\left( 1, 0, 0, 1\right)
\;\;\;\;\;\;\;\;\;\;\;\;\;\;
k^\mu_{b}\;=\; x_b\frac{\sqrt{s}}{2}\,\left( 1, 0, 0, -1\right)
\label{def3}
\;,
\end{equation}
while  the momenta of the scattered parton $c^\ast$ and of the
recoiling parton $d$, with corresponding jet rapidities $y_1$ and $y_2$
are, respectively,
\begin{equation}
k^\mu_{c^\ast}\;=\; p_\perp\,\left( \cosh y_1, \,1, 0, \sinh y_1 \right)
\;\;\;\;\;\;\;\;\;\;\;\;\;\;
k^\mu_{d}\;=\; p_\perp\,\left( \cosh y_2, -1, 0, \sinh y_2 \right)
\label{def4}
\;.
\end{equation}

$\frac{}{}$\vspace{0.5cm}
\begin{figure}
\epsfxsize=430pt
\centerline{ \epsfbox{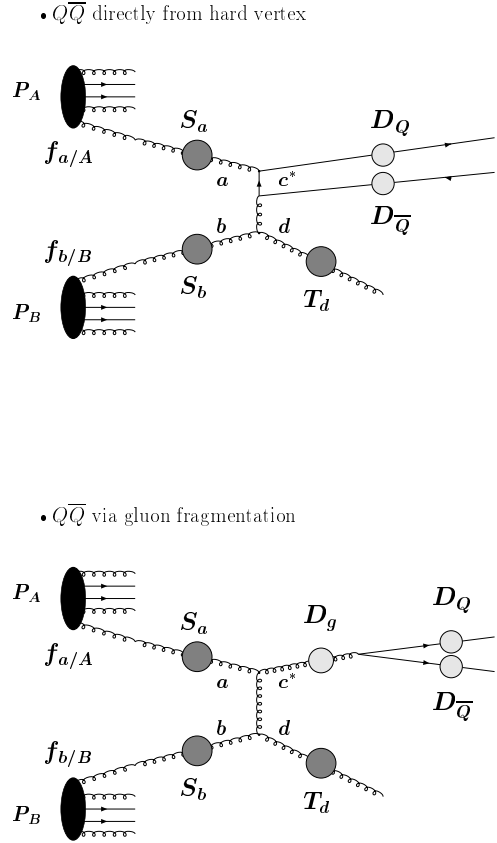} }
\vspace{-7.5cm}
\caption{
         Graphical representation of the  amplitude for the production
         and evolution of a $Q\overline{Q}$ pair, analogous to Fig. 2:
         the usual $2\rightarrow 2$ hard-scattering amplitudes for
         $ab \rightarrow c^\ast d$ are multiplied by the space-like Sudakov formfactors
         $S_{a,b}$ for the incoming partons $a$ and $b$ and 
         the outgoing parton $d$ is dressed by a time-like formfactor $T_d$, which
         contain the resummmation of real and virtual gluon emission. Similarly,
         the functions $D_g$, $D_Q$, $D_{\overline{Q}}$ embody the PQCD evolution of
         outgoing gluon-$Q\overline{Q}$ system. The initial state is weighted with 
         the parton structure-functions $f_{a/A}$ and $f_{b/B}$ of the beam particles $A,B$. 
         \label{fig:fig3}
         }
\end{figure}
\bigskip

The elementary hard process in (\ref{A1}) $ab \rightarrow c^\ast d$, where $a,b= g, q, \bar{q}$,
involves the PQCD matrix-elements for producing a $Q\bar{Q}$ with definite
color and angular momentum quantum numbers to $O(\alpha_s^3)$. These
matrix-elements have been calculated for both color-singlet \cite{CSM2,CSM3}
and color-octet \cite{Cho1} contributions for a spectrum of $Q\bar{Q}$ states.
The explicit, rather complicated expressions are well documented in the quoted
references. 
It is however useful to
plot in Fig. 4 the magnitudes and dependence with momentum transfer of the
partonic cross-sections, which are related to the squared matrix-elements as usual by
$d\hat{\sigma}/d\hat{t} = 1/(16\pi\hat{s}^2) \left| A(ab\rightarrow c^\ast d)\right|^2(\hat{s},\hat{t})$.
The Feynman diagrams that contribute to order $O(\alpha_s^3)$ 
to these partonic cross-sections are  depicted in Fig. 5.
\smallskip

\newpage
The factors $x_af_{a/A}$ and $x_bf_{b/B}$ in (\ref{A1}) give the
probablity for finding the incoming  partons $a,b$ in the beam particles $A,B$ 
with momentum fractions $x_{a,b}$. The formfactors ${\cal S}_{a,b}$
are the probabilities for the incoming partons $a,b$ to emerge
directly from the structure functions of $A,B$ without additional
initial state evolution as illustrated in Fig.3 (conversely, $1-{\cal S}_{a,b}$ 
determine the amount of real and virtual initial-state radiation). 
The explicit expression for $S_a$ (and similarly for $S_b$) reads \cite{ms44,msrep,ms3},
\begin{eqnarray}
S_a ( x_{a}, \mu^2, Q^2\,|\, t)
&=&
\exp
\left\{
\,-\, \sum_{a'}
\,
 \int_{-\infty}^t dt'
 \int d^3 \overline{r}
\,
 \int_{\mu^2}^{Q^2}
\frac{d k^{\prime\;^2}}{k^{\prime\;2}}
\,
\int_{z_-}^{z_+} \frac{d z}{z}
\,\;
\frac{\alpha_s\left((1-z) k^{\prime\;2}\right)}{2 \pi} \,
\,
\gamma_{a' \rightarrow a} (z)
\right.
\nonumber \\
& & \nonumber \\
& &
\;\;\;\;\;\;\;\;\;\;\;\;\;\;\;\;\;\;\;\;\;\;
\left.
\frac{}{}
\;\times \;
\left(
\frac{f(x_{a'}, k^{\prime \;2})} {f(x_{a}, k^{\prime\;2})}
\right)
\;\;
\theta\left(t-t'-\frac{k_0^\prime}{k^{\prime\;2}}\right)
\;
\delta^3\left(\overline{{\bf r}} - \frac{{\bf k}^\prime}{k_0^\prime}\,t'\right)
\right\}
\;,
\label{SB}
\end{eqnarray}
where the sum runs over the possible species $a' = g, q, \bar q$, and the
function $\gamma_{a'\rightarrow a}(z)$ with energy fraction
$z=k_{0\;a}/k_{0\,a'}$ is the well-known DGLAP kernel \cite{DGLAP1,DGLAP2} for the energy
distribution in the branching $a'\rightarrow a$.  The integration limits
$z^+=1-z^-=1-k^2/\mu^2$ of the $z$-integration arise from the kinematic
constraints of the branching $a\rightarrow a'$.  Last but not
least, the final-state evolution of the recoiling parton $d$ in Fig. 3
is incorporated in the time-like form factor \cite{ms44,msrep,ms3},
\begin{eqnarray}
T_d (Q^2,\mu^2\,|\,t)
&=&
\exp
\left\{
\,-\, \sum_{d'}
\,
 \int_{t}^\infty dt'
 \int d^3 \overline{r}
\,
 \int_{\mu^2}^{Q^2}
\frac{d k^{\prime\;^2}}{k^{\prime\;2}}
\,
\int_{z_-}^{z_+} d z
\;\,
\frac{\alpha_s\left((1-z) k^{\prime\;2}\right)}{2 \pi} \,
\,
\gamma_{d \rightarrow d'} (z)
\right.
\nonumber \\
& & \nonumber \\
& &
\;\;\;\;\;\;\;\;\;\;\;\;\;\;\;\;\;\;\;\;\;\;
\;\;\;\;\;\;\;\;\;\;\;\;\;\;\;\;\;\;\;\;\;\;
\left.
\frac{}{}
\;\times \;
\theta\left(t-t'-\frac{k_0^\prime}{k^{\prime\;2}}\right)
\;
\delta^3\left(\overline{{\bf r}} - \frac{{\bf k}^\prime}{k_0^\prime}\,t'\right)
\right\}
\;,
\label{TB}
\end{eqnarray}
which is summed over the species $d'= g , q, \bar q$
of  the possible daughter species $d'$, and the phase-space boundary
for the branching is again $z^+=1-z^-=1-\mu^2/k^2$.
\bigskip

I emphasize that the expressions ({\ref{SB}) and (\ref{TB}) are `space-time
generalizations' of the standard space-like and time-like form factors with
so-called angular ordering derived within the leading-log approximation
including coherence effects \cite{DGLAP2,MLLA}.  By integrating over all times
$t\rightarrow \infty$ and over the spatial variables, one would regain the
commonly used momentum-space expressions that make no reference to the
space-time development of the evolution.  Characteristic of the above
space-time dependent form factors is the appearence of the $\theta$- and
$\delta$-functions: First, the factor $\theta(\delta t-k_0/k^2)$ accounts for
the mean formation time $t_{form} = \gamma \tau_{form} = (k_0/k)(1/k)$ of 
a gluon emission with the energy  $k_0$. By enforcing that $t_{form}$ must
be smaller than or equal to the considered time interval $\Delta t = t - t'$,
one ensures that the emitted offspring
becomes incoherently separate from radiating mother. Notice that instead of 
$\theta(\Delta t-t_{form})$, one could use a smeared-out distribution such
as $\exp(-\Delta t/t_{form})$.  The results are rather independent of the
specific form, provided only  that the choice respects the uncertainty 
principle.  Secondly, the factor $\delta^3\left(\overline{{\bf
r}} - {\bf v}\,t'\right)$ with the velocity ${\bf v} = {\bf k}/k_0$ simply
reflects the classical propagation of each parton in the cascade evolution,
i.e. the spatial development is approximated by straight-line
trajectories of the partons between branching vertices.
\newpage
$\frac{}{}$\vspace{1.5cm}
\begin{figure}
\epsfxsize=430pt
\centerline{ \epsfbox{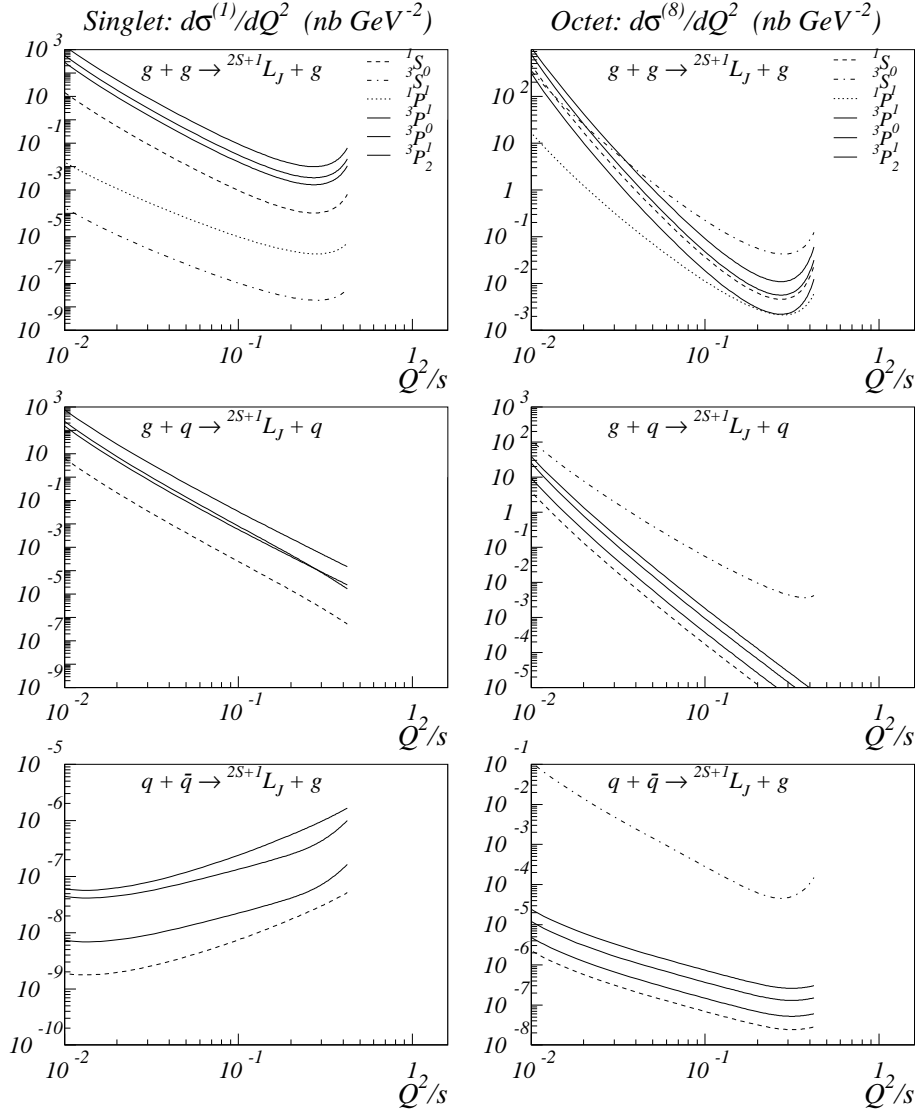} }
\vspace{-3.0cm}
\caption{
         $O(\alpha_s^3)$ PQCD cross-sections for the production of  $Q\overline{Q}$
         states with specific quantum numbers $J,L,S$ by a hard parton collisions with
         energy $\sqrt{\hat{s}}$ and momentum transfer $Q=p_\perp$, 
         mediated by $gg$, $qg$, and $q\bar{q}$ collisions. Notice the very different 
         magnitude of color-singlet (left) and color-octet cross-sections (right).
         \label{fig:fig4}
         }
\end{figure}
\newpage
$\frac{}{}$\vspace{1.5cm}
\begin{figure}
\epsfxsize=430pt
\centerline{ \epsfbox{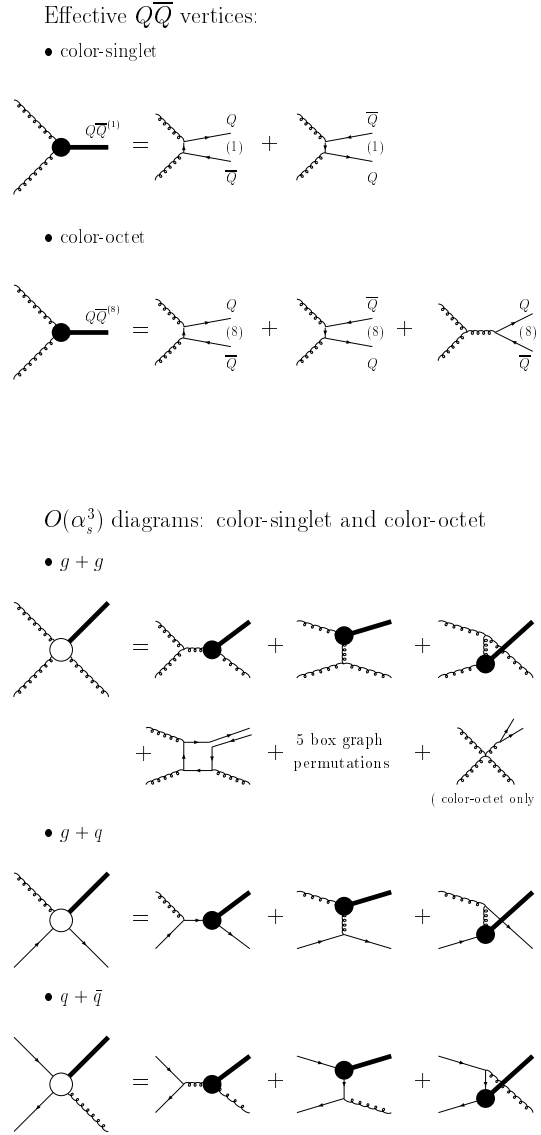} }
\vspace{-1.5cm}
\caption{
         Color-singlet and color-octet Feynman diagrams that contribute at  
         $O(\alpha_s^3)$ to $Q\overline{Q}$ production via the processes
         (i) $g+g\rightarrow Q\overline{Q} +g$, (ii) $g+q\rightarrow Q\overline{Q} +q$,
         and (iii) $q+\bar{q}\rightarrow Q\overline{Q} +g$. The top part defines the solid
         blobs with a fat line attached as effective vertices, which are
         to be understood as insertions in the diagrams of the lower part.
         \label{fig:fig5}
         }
\end{figure}
\newpage

\subsection{Evolution}
\medskip

The amplitude
$
{\cal A}\left(c^\ast \rightarrow \left[Q\bar{Q}\right]^{JLS, \,(1)} + m\;g 
+ m'\;q+ \overline{m}'\bar{q} \right)_{evolution}
$
describes the perturbative evolution subsequent to the hard production of the
time-like fluctuation $c^\ast=Q\bar{Q}$ or   $c^\ast = g$.
If $c^\ast = Q\bar{Q}$ (Fig. 2a, 3a), the $Q\bar{Q}$ pair is produced
directly at the hard vertex, i.e., without intermediate gluon, so that there
is only one stage, namely the evolution of the $Q\bar{Q}$ until it
becomes a hadron $H_{Q\bar{Q}}$.  
If $c^\ast = g$
(Fig. 2b, 3b), there are  two stages involved: first, the evolution of the time-like
gluon through real and virtual gluon radiation until it decays into a
(off-shell) $Q\bar{Q}$ pair, and second the analogous evolution of the
virtual $Q\bar{Q}$ until it converts into a bound-state upon hadronization.
Both cases can incorporated by
representing ${\cal E}^2_{c^\ast}$ as follows:
\begin{eqnarray}
{\cal E}^2_{c^\ast}\left(Q^2, \mu^2 
\,|\, k_1, k_2, \overline{{\bf r}}_1,\overline{{\bf r}}_2, t \right)
&\equiv&
\left|
{\cal A}\left(c^\ast \rightarrow \left[Q\bar{Q}\right]^{JLS, \,(1)} + m\;g 
+ m'\;q+ \overline{m}'\bar{q} \right)_{evolution}
\right|^2
\nonumber \\
& &
\nonumber \\
&=&
\int_{t_0}^{t} 
\int d^3 \overline{r}_{c^\ast}
\;
\int_{\mu^2}^{Q^2}\frac{dk^2}{k^2} \int dx\int dz \; 
{\cal F}_{c^\ast}\left(x,z,Q^2,k^2\,|\,\overline{{\bf r}}_{c^\ast}, t\right)
\nonumber \\
& &
\nonumber \\
&&
\;\times \;
\frac{1}{zx}\,
D_{Q}^{Q}\left(\frac{x_1}{zx},k^2,\mu^2\,|\,\overline{{\bf r}}_1, t\right)
\;
\frac{1}{(1-z)x}\,
D_{\overline{Q}}^{\overline{Q}}\left(\frac{x_2}{(1-z)x},k^2,\mu^2
\,|\,\overline{{\bf r}}_2, t\right)
\label{A2}
\;.
\end{eqnarray}
Here the space-time dependent parton densities 
$D_{a}^{b}(x,k^2,\mu^2\,|\,\overline{{\bf r}}, t)$ embody
the resummation  of the leading-log
corrections from virtual and real gluon emsisions to all orders in $\alpha_s$ 
according to the DGLAP equations \cite{DGLAP1}, that is,
\begin{equation}
D_{a}^{b}(x,k^2,\mu^2\,|\,\overline{{\bf r}}, t)
\;=\;
\sum_{c}
\,
 \int_{t_0}^t dt'
 \int d^3 \overline{r}
\,
 \int_{\mu^2}^{k^2}
\frac{d k^{\prime\;^2}}{k^{\prime\;2}}
\,
\int d z
\frac{\alpha_s\left((1-z) k^{\prime\;2}\right)}{2 \pi} \,
\,
\gamma_{a \rightarrow c} (z)
\;\,
\theta\left(t-t'-\frac{k_0^\prime}{k^{\prime\;2}}\right)
\;
\delta^3\left(\overline{{\bf r}} - \frac{{\bf k}^\prime}{k_0^\prime}\,t'\right)
\label{D1}
\;,
\end{equation}
with $\gamma_{a\rightarrow c}(z)$ again denoting the  DGLAP
kernel for branching $a\rightarrow c$.
In (\ref{A2}) the function
${\cal F}_{c^\ast}$ depends on whether the $Q\bar{Q}$ pair is
produced directly at the hard vertex (fig. 2a, 3a)
or  via gluon fragmentation (Fig. 2b, 3b):
\begin{equation}
{\cal F}_{c^\ast}\left(x,z,Q^2,k^2\,|\,\overline{{\bf r}}_{c^\ast}, t\right)
\;=\;
\left\{
\begin{array}{ll}
\delta(1\,-\,x)\;\,\delta\left(1\,-\,\frac{k^2}{Q^2}\right)
\;\delta(\overline{{\bf r}}_{c^\ast}\,-\,\overline{{\bf R}}_0)
& \;\;\;\;\; \mbox{if} \;\; c^\ast\;=\; Q\bar{Q}
\\
D_{g}^{g}\left(x,Q^2,k^2\,|\,\overline{{\bf r}}_{c^\ast}, t\right)
\;\frac{\alpha_s((1-z)k^2)}{2\pi}\,\Gamma_{g\rightarrow Q\bar{Q}}(z)
& \;\;\;\;\; \mbox{if} \;\; c^\ast\;=\; g
\end{array}
\right.
\label{F}
\;.
\end{equation}
On the right side,
for the case $c^\ast = g$, the kernel
$\Gamma(z)$ is now the {\it real} branching probability
and not the sum of real and virtual contributions as the above $\gamma(z)$,
since  a {\it real} $Q\bar{Q}$ pair has to be produced.
The two functions are related, however, by
\begin{equation}
\gamma_{a\rightarrow b}(z) \;=\; 
\Gamma_{a\rightarrow bb'}(z) \;-\; 
\delta_{ab}\,\delta(z-1)\;\frac{1}{2}\sum_{c,c'}\int_0^1dy 
\;\Gamma_{a\rightarrow cc'}(y)
\;.
\end{equation}
For the case $c^\ast = Q\bar{Q}$, the product of $\delta$-functions
simply cause the integrations over the intermediate $c^\ast$ to collapse
and enforce the $Q\bar{Q}$ to be produced directly at the
mean position $\overline{{\bf R}}$ at the hard vertex at time $t_0$.
\bigskip

The {\it physical interpretation} of eq. (\ref{A2}) in conjunction with
(\ref{F}) is rather evident:  the product of the functions 
$D$ contain the cumulative effect of
the perturbative evolution of the produced $c^\ast$ at $Q^2$
and of the final $Q\bar{Q}$ pair:
either
(i)
directly,
if $c^\ast = Q\bar{Q}$, the  evolution of the $Q\bar{Q}$ from $Q^2$ to
the point when it hadronizes, but at most to the factorization point $\mu^2$
in (\ref{def1});
or (ii)
by fragmentation of the gluon, if $c^\ast = g$, from $Q^2$ to $k^2$, the decay into
$Q\bar{Q}$ at $k^2$  and the subsequent evolution of the pair 
from $k^2$ to $\mu^2$ (or above $\mu^2$, if the $Q\bar{Q}$ form a bound-state
beforehand, in which case the evolution terminates).
\bigskip
\bigskip
\bigskip

\subsection{Coalescence}
\medskip

The final ingredient in the total amplitude (\ref{ansatz}),
is the subamplitude 
${\cal A}\left( \left[Q\bar{Q}\right]^{JLS,(1)} \rightarrow 
H_{Q\bar{Q}}(^{2S+1}L_J)\right)_{coalescence}$
for the conversion
of the final state $Q\bar{Q}$ pair into the bound-state hadron  $H_{Q\bar{Q}}$.
Again, I am interested in the square of this subamplitude,
\begin{equation}
{\cal C}^2_{JLS}
\left(\mu^2 \,|\, k_1, k_2, \overline{{\bf r}}_1,\overline{{\bf r}}_2, t \right)
\;\equiv\;
\left|
{\cal A}\left( \left[Q\bar{Q}\right]^{JLS,(1)} \rightarrow 
H_{Q\bar{Q}}(^{2S+1}L_J)\right)_{coalescence}
\right|^2
\;.
\label{A3}
\end{equation}
Clearly, in view of the unknown details of the non-perturbative physics, the
coalescence of $Q\overline{Q}$ to a bound-state and eventually to a physical
quarkonium state, requires some model input, and therefore introduces 
a model dependence contrary to the previous
elements of the PQCD description of production and evolution.
The rationale is here, to avoid the introduction of free parameters 
and develop a transparent and practical bound-state formalism
within the principles of NRQCD.
\bigskip

\subsubsection{Scheme of approximation}

Starting from the amplitude level,
the `unsquared' ${\cal C}$ can be expressed in the usual fashion
of time-dependent perturbation theory on the basis of the Dyson series \cite{Sakurai}, 
\begin{eqnarray}
{\cal C}_{JLS}
\left(\mu^2 \,|\, k_1, k_2, \overline{{\bf r}}_1,\overline{{\bf r}}_2, t \right)
&=&
-i\int_{t_0}^t dt'\;
\langle \;\Psi\;\vert\;V(\mu^2,\overline{{\bf R}},t')\;\vert\;\left[\psi\,\overline{\psi}\right]\;\rangle 
\nonumber \\
& & \nonumber \\
&=&
\langle \;\Psi\;\vert\;\widetilde{V}(\mu^2,\overline{{\bf R}})\;\vert\;\left[\psi\,\overline{\psi}\right]\;\rangle 
\;\,\left( 1\,-\,e^{i\,\Delta E \, t}\right)
\label{A3a}
\;,
\end{eqnarray}
where $\Delta E = E_{Q\bar{Q}} - E_H$, 
the wave-functions of the $Q$ and $\bar{Q}$ are
denoted by $\psi$, $\overline{\psi}$
(the brackets $[\ldots]$ indicating the  $Q\bar{Q}$ coupling to a color-singlet 
and specific total spin state as described
below),
and $\Psi$ is the bound-state wave-function of $H_{Q\bar{Q}}$.
The transition operator 
\begin{equation}
V(\mu^2,\overline{{\bf R}},t)\;=\;
\widetilde{V}(\mu^2,\overline{{\bf R}}) \; \,e^{i\,\Delta E \, t}
\label{V}
\end{equation}
represents 
the non-perturbative mechanism of conversion from the partonic
$Q\bar{Q}$ state to the hadronic bound-state $H_{Q\bar{Q}}$,
which in principle may depend on time $t$, the center-of-mass
position $\overline{{\bf R}}$,  and of course
on the factorization scale $\mu^2$.
Unfortunately the non-peturbative operator $V$ cannot inferred from first principles
at present, so that one is forced to rely on lattice results, or
phenomenological model building. 
Most previous quarkonium studies  simply treated the 
unknown non-perturbative amplitude in (\ref{A3}) or (\ref{A3a}) as a parameter, 
i.e., a number being
determined either from fits to experimental data or
taken from NRQCD potential models.
\smallskip

Instead, I follow  a different approach and
 try to actually {\it calculate} approximately this
amplitude on the basis of the following arguments:
It is well known from hadronization
studies of jet fragmentation that the mechanism
of parton-hadron conversion at high energies appears to be a 
universal and local
phenomenon \cite{LPHD1,LPHD2}, in the sense that it is rather independent
of the space-time history of the hadronizing partons and
essentially determined by the final-state configuration
of neighboring partons in a local phase-space cell.
Moreover it is known that that the actual details
of the hadronization dynamics are irrelevant for
simple observables as  hadron production rates.
Hence, it is near at hand to assume that the operator
$V$, eq. (\ref{V}), is approximately of local nature, in the sense 
that its time variation is sufficiently slow that it may be omitted
with respect to the $Q\bar{Q}$ motion,
and furthermore is at best weakly dependent on $\overline{{\bf R}}$
(which in the presence of translational invariance  is certainly the case).
Accordingly, I make the following key approximation, pulling $\widetilde{V}$ out 
of the matrix-element (\ref{A3a}):
\begin{equation}
\langle \;\Psi\;\vert\;\widetilde{V}(\mu^2,\overline{{\bf R}})\;\vert\;\left[\psi\,\overline{\psi}\right]\;\rangle 
\;\;
\longrightarrow\;\;
\langle \widetilde{V}\rangle \;
\times \;
\langle \;\Psi\;\vert\;\left[\psi\,\overline{\psi}\right]\;\rangle 
\label{Vapprox}
\;.
\end{equation}
Returning to (\ref{A3a}), I insert (\ref{Vapprox}), take the square,
and integrate over the hadronic energy spectrum $dE_H$ weighted with the level density 
$\rho(E_h)$, and obtain so the approximate result for the
squared amplitude ${\cal C}^2$ of eq. (\ref{A3}):
\begin{eqnarray}
{\cal C}^2_{JLS}
\left(\mu^2 \,|\, k_1, k_2, \overline{{\bf r}}_1,\overline{{\bf r}}_2, t \right)
&\approx&
\int_\mu dE_H \,\;\rho(E_H)
\;
\left|\langle \widetilde{V}\rangle\right|^2
\;\frac{4}{(\Delta E)^2}\;\sin^2\left(\frac{\Delta E\,t}{2}\right) \;\;
\left|\langle \;\Psi\;\vert\;\left[\psi\,\overline{\psi}\right]\;\rangle\right|^2 
\nonumber \\
& & \nonumber \\
&\stackrel{t\rightarrow \infty}{=}&
\;\;\;\;\;\;\;\;
2\pi\,t\;\rho(E_H)
\;
\left|\langle \widetilde{V}\rangle\right|^2
\left|\langle \;\Psi\;\vert\;\left[\psi\,\overline{\psi}\right]\;\rangle\right|^2 
\nonumber \\
& & \nonumber \\
&\simeq&
\;\;\;\;\;\;\;\;
\left|\langle \;\Psi\;\vert\;\left[\psi\,\overline{\psi}\right]\;\rangle\right|^2 
\;,
\label{A3b}
\end{eqnarray}
Here I have in the second step  exploited the fact that only those final states
contribute significantly for which $t\sim 2\pi/\Delta E$, i.e., 
$\Delta t \Delta E \sim 1$ with $\Delta t = t-t_0$, and
in the last step I have assumed that the energy spectrum $\rho(E_H)\propto 1/E_H$,
which is roughly satisfied by the known hadron states.
Because the
time scale $\Delta t$ for the conversion from $Q\bar{Q}$ to $H_{Q\bar{Q}}$
is related by the
uncertainty principle to the energy difference $\Delta E$
between the quarkonic and hadronic states, 
the  approximation (\ref{Vapprox}) and (\ref{A3b}) reduces the
unknown matrix-element (\ref{A3}) to the
simple overlap matrixelement between the heavy
quark pair $Q\bar{Q}$ and the  bound-state hadron $H_{Q\bar{Q}}$.
\bigskip

\subsubsection{Wave-functions}

To proceed from (\ref{A3b}), the goal is  to model the  non-perturbative 
dynamics of conversion of the final $Q$ and $\bar{Q}$ 
into the bound-state hadron $H_{Q\bar{Q}}$ by making a specific
ansatz for the wave-functions $\Psi$ and $\psi,\overline{\psi}$.
I shall adopt the
well-established NRQCD bound-state picture for quarkonia 
\cite{CSM1} in conjuction
with coalescence framework developed in Refs. \cite{coalescence}.
As proclaimed before, I restrict the considerations to
$S$-wave and $P$-wave bound-states with angular momentum $L=0$
and $L=1$, respectively.
The corresponding bound-state wave-functions are therefore
of scalar, respectively vector character in angular
momentum space.
In the  quarkonium bound-state formalism, developed in
detail in Ref. \cite{CSM1}, the bound-state wave-function 
is constructed in terms of the
overlap amplitude between quark and antiquark spinors
$\psi$ and $\overline{\psi}$ in a given
total spin configuration (singlet $S=0$ or triplet $S=1$) 
and the bound-state wave-function $\Psi$ in a given angular
momentum state $L=0$ or $L=1$, to form s state of total
angular momentum $J=0,1,2$.
As usual, I assume that in the restframe of the bound-state
the relative momentum $|\bf{k}|=\frac{1}{2}|\bf{k}_1-\bf{k}_2|$
of the $Q$ and $\bar{Q}$ with 3-momenta $\bf{k}_1$ and $\bf{k}_2$, 
respectively, is small as compared to the quark mass $M_Q$. 
As a consequence, one can decompose 
the (unsquared) amplitude ${\cal C}_{JLS}$, eq. (\ref{A3b}),
in coordinate space as follows:
\begin{equation}
{\cal C}_{JLS}
\left(\mu^2 \,|\, k_1, k_2, \overline{{\bf r}}_1,\overline{{\bf r}}_2, t \right)
\;=\;
2\pi \delta\left(t-\frac{t_1+t_2}{2}\right) \;
\;\frac{1}{\sqrt{2J+1}}\;
\sum_{M,S_z} \;\langle\;L \,M,\;S\, S_z\,| \,J \,J_z\;\rangle
\;\; \Psi_M^{L\;\ast}
\left({\bf r}_1,{\bf r}_2\right)
\;
\left[\psi\,\overline{\psi}\right]_{S_z}^{S \,(1)} \left({\bf r}_1,{\bf r}_2\right)
\;.
\label{chi1} 
\end{equation} 
Here $\langle\;L \,M,\;S\, S_z\,| \,J \,J_z\;\rangle$ is the Clebsch-Gordan coefficient
for the coupling of spin $S$ and relative angular momentum $L$ to a total
$J$ with associated projections $S_z$, $M$,and $J_z$,
respectively, $\Psi_M^{L\;\ast}$ is the complex conjugate of the bound-state
wave-function for a $L$-state with projection $M$, and
\begin{equation} 
\left[\psi\,\overline{\psi}\right]_{S_z}^{S\;(1)}({\bf r}_1,k_1; {\bf r}_2,k_2)
\;=\;
\;\frac{1}{\sqrt{2\,N_c}}\;
\sum_{c_1,c_2=1}^{N_c} \;\langle\;3 \,i,\;\overline{3}\, j\,| \,1 \,0\;\rangle
\sum_{s_1,s_2} \;\langle\;\frac{1}{2} \,s_1,\;\frac{1}{2}\, s_2\,| \,S \,S_z\;\rangle
\;\,
\psi\left({\bf r}_1, s_1, c_1\right)\; \overline{\psi}\left({\bf r}_2, s_2, c_2\right)
\label{chi2} 
\end{equation} 
couples the $Q$ and $\overline{Q}$ spinors to a color-singlet with total spin
$S$ and projection $S_z$.

To proceed, one needs to specify  the wave-functions $\Psi$ and $\psi$, $\overline{\psi}$.
Most plausible would be to use quarkonium wave functions from potential model
calculations, which however are not available analytic form.
Therefore, I choose instead to work with harmonic oscillator wave-functions
for which it is possible to obtain the overlap integrals involving
$\Psi, \psi, \overline{\psi}$,  as transparent, closed expressions.
Following the coalescence aproach of Refs. \cite{coalescence},
I shall assume that  quark $Q$ and antiquark $\bar{Q}$ may be described by
wave-packets of width $\sigma_i \propto 1/\sqrt{k_i^2}$
($k_i^2=k_{0\,i}^2-{\bf k}_i^2$), localized in space around
${\bf \overline{r}}_i$ and in momentum space around
${\bf \overline{k}}_i$, multiplied by the usual $SU(2)$ and $SU(3)$ spinors 
$\chi$, $\xi$ for spin and color, respectively:
\begin{equation}
\psi\left({\bf r}_i,s_i, c_i\right)\;=\;
\frac{1}{(\pi\,\sigma_i^2)^{3/4}}\;
\exp\left(-\frac{({\bf r}_i-{\bf  \overline{r}}_i)^2}{2 \,\sigma_i^2} \right)
\;\,\exp(i\,{\bf \overline{k}}_i\cdot {\bf r}_i)
\;\;\chi^{\frac{1}{2}}_{s_i}\;\;\xi^{\frac{1}{3}}_{c_i}
\;\;\;\;\;\;\;\;\;
i = 1,2 \;\;(=Q,\,\bar{Q})
\label{psi}
\end{equation}
Here and in the following the notation of 3-vectors is such that, e.g.,  
${\bf \overline{r}}_i$  refers to the mean position as a classical
variable, whereas ${\bf r}_i$ describes the quantum-mechanical fluctuation
of a particle's position around the  mean value.
Notice that the plane-wave factor $\exp(i\,{\bf \overline{k}}_i\cdot {\bf r}_i)$
ensures the uncertainty principle in the sense that a classical particle interpretation
is only possible if  $|{\bf r}| << 1/|{\bf \overline{k}}_i|$.

The bound-state wave function $\Psi$ on the other hand, I write as a
product of its center-of-mass motion and its internal motion, again in terms
of Gaussian wave-packets, the size of which generally depends on the 
particular orbital angular momentum state, and of course on the flavor of
the heavy-quark pair.
Defining center-of-mass and relative coordinates and momenta as
\begin{eqnarray}
{\bf R} &=& \frac{{\bf r}_1 \,+\, {\bf r}_2}{2}
\;\;\;\;\;\;\;\;\;\;\;\;\;\;\;
{\bf r} \;=\;  {\bf r}_1 \,-\, {\bf r}_2 
\nonumber \\
{\bf K} &=&  {\bf k}_1 \,+\, {\bf k}_2
\;\;\;\;\;\;\;\;\;\;\;\;\;\;\;
{\bf k} \;=\; \frac{ {\bf k}_1 \,-\, {\bf k}_2 }{2}
\;,
\label{def5}
\end{eqnarray}
and similarly for the mean positions ${\bf \overline{R}}$, ${\bf \overline{r}}$
and momenta  ${\bf \overline{K}}$, ${\bf \overline{k}}$,
I make the factorized ansatz
\begin{equation}
\Psi^L_M({\bf r}_1, {\bf r}_2) \;=\;
\Phi^L({\bf R}) \;\,\phi^L_M({\bf r})
\label{Psi1}
\;.
\end{equation}
The center-of-mass wave-function $\Phi^L$ depends on $L$ in its spatial width,
\begin{equation}
\Phi^L({\bf R}) \;=\;
\frac{1}{(\pi\,\Sigma_L^2)^{3/4}}\;
\exp\left(-\frac{({\bf R}-{\bf  \overline{R}})^2}{2 \,\Sigma_L^2} \right)
\;\,\exp(i\,{\bf \overline{K}}\cdot {\bf R})
\end{equation}
and the relative motion $\phi^L_M$ depends on both $L$ and its projection $M$,
\begin{equation}
\phi^L_M({\bf r})
\;=\;
\frac{{\cal N}_L}{(\pi\,\alpha_L^2)^{3/4}}\;\,
r^L\;\,
\exp\left(-\frac{r^2}{2 \,\alpha_L^2} \right)
\;\;Y_{LM}({\bf \widehat{r}})
\label{Psi2}
\;,
\end{equation}
with the $L$-dependent prefactors  ${\cal N_0} =1$ and ${\cal N}_1=\sqrt{2/3}$ and 
the spherical harmonics  $Y_{LM}$ depending on the angular orientation
${\bf \widehat{r}}={\bf r}/r$, where $r \equiv |{\bf r}|$.
\bigskip

\subsubsection{Coalescence probabilities}

Inserting the wave-functions $\psi,\overline{\psi}$ and $\Psi$ into
(\ref{chi2}),
integrating over ${\bf r}_1={\bf R}+\frac{1}{2}{\bf r}$ and
${\bf r}_1={\bf R}-\frac{1}{2}{\bf r}$, 
summing over the angular momentum projections,
and then squaring the resulting amplitude ${\cal C}_{JLS}$,
one obtains
the  {\it coalescence  probability} ${\cal C}^2_{JLS}$  for  the 
conversion of $Q\bar{Q}$ into bound-states $^{2S+1}L_J$.
Specifically for the $S$- and $P$-wave states,  one obtains
the following closed expressions:
\begin{itemize}
\item
$S$-states ($L=0$, $J=S=0,1$):
\begin{equation}
{\cal C}^2_{J0S}
\left(\mu^2 \,|\, \overline{k}^2,\overline{r}^2, t \right)
\;=\;
\;\left(\frac{1}{4\pi}\right)\;\; 
\left(\frac{4 \,\nu_0}{\sqrt{2}\,\mu_0}\right)^3
\;\; \exp\left( - \frac{\nu_0^2}{2}\;\overline{k}^2\right) 
\;\; \exp\left( - \frac{1}{\mu_0^2}\;\overline{r}^2\right)
\label{CJ0S}
\end{equation}
\item
$P$-states ($L=1$, $J=L+S=0,1,2$):
\begin{equation}
{\cal C}^2_{J1S}
\left(\mu^2 \,|\, \overline{k}^2,\overline{r}^2, t \right)
\;=\;
\;\left(\frac{3}{4\pi}\right)\;\; 
\left(\frac{4 \,\nu_1}{\sqrt{2}\,\mu_1}\right)^3\; 
\;\;\frac{1}{6}
\, \left(\frac{\alpha_1^4}{\mu_1^4}\;\overline{r}^2\;+\; \nu_1^4\;\overline{k}^2\right)
\;\; \exp\left( - \frac{\nu_1^2}{2}\;\overline{k}^2\right) 
\;\; \exp\left( - \frac{1}{\mu_1^2}\;\overline{r}^2\right)
\label{CJ1S}
\;,
\end{equation}
\end{itemize}
In the above formulae (\ref{CJ0S}) and (\ref{CJ1S})
I used $\overline{k} = |{\bf \overline{k}}| = \frac{1}{2}|{\bf \overline{k}}_1 -{\bf \overline{k}}_2|$,
$\overline{r} = |{\bf \overline{r}}| = |{\bf \overline{r}}_1 -{\bf \overline{r}}_2|$,
and
\begin{equation}
\mu_L^2\;\equiv\; 2\,\sigma^2\;+\alpha_L^2
\;\;\;\;\;\;\;\;\;\;\;\;
\nu_L^2\;\equiv\; \frac{\sigma^2\;\alpha_L^2}{\mu_L^2}
\;.
\label{munu}
\end{equation}
Furthermore, $\sigma^2 = \frac{1}{2}(\sigma_1^2+\sigma_2^2)$, and
$\Sigma_L^2 = \sigma^2/2$, assuming that the 2-body center-of-mass motion
of $Q$ and $\overline{Q}$ is unaltered during the coalescence process.
\smallskip

It is worth noting that the final expressions for the coalescence probabilities 
${\cal C}^2_{JLS}$ in (\ref{CJ0S}) and (\ref{CJ1S}) are functions
of the relative $Q\overline{Q}$ momentum and separation only, and do not
depend neither on the individual variables ${\bf \overline{k}}_i$ 
and ${\bf \overline{r}}_i$, nor on the total momentum  ${\bf \overline{K}}$ 
or the $cm$-position ${\bf \overline{R}}$ of the pair.
This makes the coalescence process a {\it local} mechanism, in accordance with
the aforementioned locality of hadronization in high-energy particle collisions.
Also notice that the coalescence probabilities are time-dependent, although
not explicit in (\ref{CJ0S}) and (\ref{CJ1S}), due to the dynamically
changing $Q$ and $\overline{Q}$ momenta and trajectories, i.e.,
$\overline{k}= \overline{k}(t)$ and $\overline{r}= \overline{r}(t)$.
Finally, the dependence on $\mu^2$ indicated by the first argument 
of ${\cal C}^2_{JLS}$, is again implicit in the $Q$ and $\overline{Q}$
momenta, since per definition $\mu^2$ serves as a cut-off for
soft gluon radiation with energies below $\mu$ from $Q$ or $\overline{Q}$
and hence limits the radiative energy loss of the pair. This in turn
influences the coalescence to a $Q\overline{Q}$ bound-state, although
this dependence is very weak.
\medskip

The independent {\it parameters} involved in (\ref{CJ0S}) and (\ref{CJ1S})
are $\sigma^2$ (a property of the $Q$ and $\overline{Q}$ upon coalescence,
but independent of the bound-states $H_{Q\overline{Q}}$), and
$\alpha_L^2$ (a property of the bound-states $H_{Q\overline{Q}}$,
depending on the angular momentum quantum number $L$ as well as on
the `flavor', $c\bar{c}$ or $b\bar{b}$).
However, these parameters $\sigma^2$ and $\alpha_L^2$ are {\it not for free choice}.
Rather than that, the possible values of $\sigma^2$,
are given by the invariant (off-shell) masses $k_i^2=k_{0\,i}^2-{\bf k}_i^2$, 
and hence are dictated
by the dynamics of $Q$ and $\overline{Q}$ itself, whereas $\alpha_L^2$ is
determined by the Bohr radii of the various quarkonium states
$H_{Q\overline{Q}}(^{2S+1}L_J)$.
Specifically, for 
$\sigma^2 = \frac{1}{2}(\sigma_1^2+\sigma_2^2)$, 
the widths $\sigma_i$ of the $Q$ and $\overline{Q}$ wave-packets are obtained
from relating their (time-evolving)
invariant masses $k_i^2$ by means of the uncertainty principle 
to the corresponding mean spatial sizes
\footnote{
The factors 2/3 in (\ref{param1}) and (\ref{param2}) arise
from the definition of the mean square radii of the heavy quarks
and the bound-state, respectively, which involve the ratios 
$\langle r^2 \rangle = \int d^3r' \,r^{'\,2}\, f(r')/ \int d^3r' \,f(r')$,
where $f(r)$ stands for the Gaussian wave-packets of $Q$, $\overline{Q}$ and 
of the bound-state.
},
\begin{equation}
\sigma_i^2 \;=\; \frac{2}{3\,k_i^2}
\;,\;\;\;\;\;\;\;\;\;\;
k_i^2 = k_{0\,i}^2 - {\bf k}_i^2 \,\ge\, M_Q^2 
\;, 
\label{param1}
\end{equation}
whereas the widths $\alpha_L$  of the  bound-state wave-functions
are related to the sizes of the corresponding quarkonium hadrons
$H_{Q\overline{Q}}(^{2S+1}L_J)$ through the measurable electron scattering
radii,
\begin{equation}
\alpha_L^2 \;\equiv\; \alpha_L^2 (H_{Q\overline{Q}})
\;=\;\frac{2}{3}\; R_{Q\overline{Q}}^2
\;,\;\;\;\;\;\;\;\;\;\;
R_{Q\overline{Q}}\;\equiv\; R_{Q\overline{Q}}(^{2S+1}L_J)
\;.
\label{param2}
\end{equation}
The relevant values of $R_{Q\overline{Q}}$
for charmonium and bottonium states that are used later, are listed
in Table 1.
\vspace{-1.0cm}
\begin{center}
\begin{minipage}[t]{15.0cm}
\begin{table}[t]
\epsfxsize=425pt
\centerline{ \epsfbox{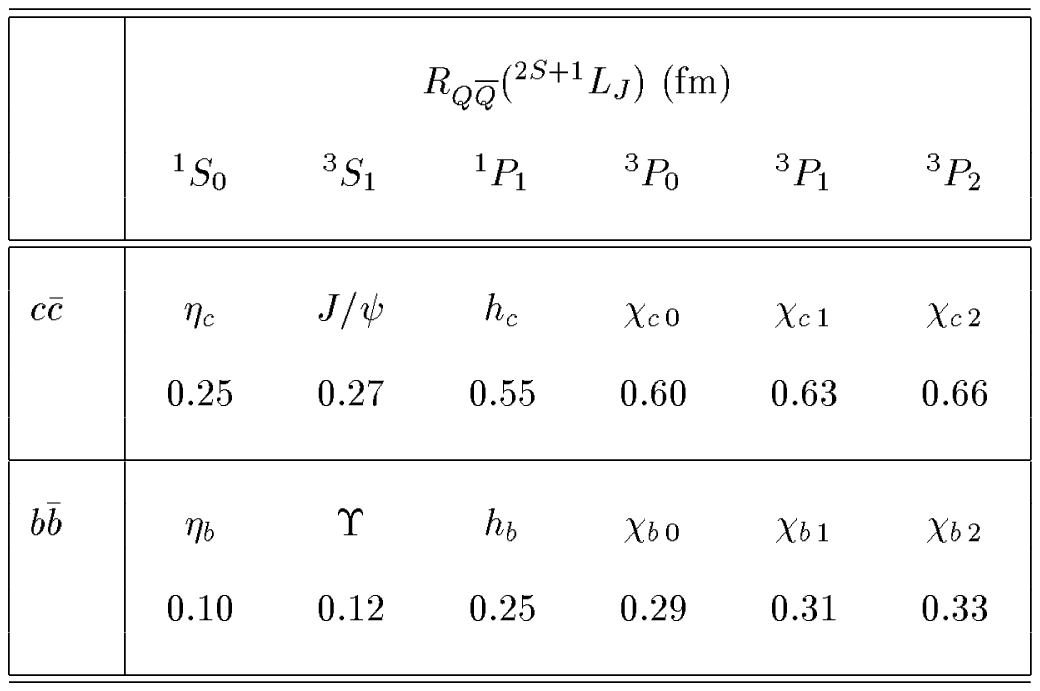} }
\vspace{-12.5cm}
\caption{Quarkonium RMS values $R_{Q\overline{Q}}$ used for the widths of 
         bound-state wave functions with different $J,L,S$ in conjunction
         with the relation (\ref{param2}).
         }
\end{table}
\end{minipage}
\end{center}

The coalescence probabilities (\ref{CJ0S}) and (\ref{CJ1S}) consist (apart from the
normalization factors that are always $\le 1$) of two Gaussians, one in momentum space,
$\exp( - \nu^2\overline{k}^2/2)$, and one in position space, $ \exp( -\overline{r}^2/\mu^2)$.
The first one ensures the conservation of momentum, whereas the second restricts
the distance between the coalescing $Q$ and $\bar{Q}$ to a width which is given by the
bound-state size, augmented by the uncertainty in the positions of the $Q$ and $\bar{Q}$.
The essential difference between the $S$-wave coalescence probabilities (\ref{CJ0S}) and 
the $P$-wave counterpart  (\ref{CJ1S}), is the factor
$(\alpha^4\overline{r}^2/\mu^4+ \nu^4\overline{k}^2)$ present in the latter.
For ${\bf \overline{k}}\rightarrow 0$ the $P$-wave probability ${\cal C}^2_{J1S}$ therefore
vanishes, reaches a maximum as ${\bf \overline{r}}$ increases, and then
tends to zero again with an exponential tail.
Contrary to this,
the $S$-wave probability ${\cal C}^2_{J0S}$,
for ${\bf \overline{k}}\rightarrow 0$, has a maximum
at ${\bf \overline{r}} = 0$ and damps out exponentially as  ${\bf \overline{r}}$ increases.
This characteristic behavior is plotted in Fig. 6, which shows
the coalesence probabilities  (\ref{CJ0S}) and (\ref{CJ1S}) for the different
angular momentum eigenstates $^{2S+1}L_J$ as a function of $\overline{r}\equiv|{\bf \overline{r}}|$ 
for various values of $\overline{k}\equiv|{\bf \overline{k}}|$. 
\bigskip

\subsubsection{Relation between  ${\cal C}_{JLS}$ and commonly used 
quarkonium wave-functions ${\cal R}_{L}(0)$}

At this point, it is instructive to relate the coalescence probabilities
${\cal C}^2_{JLS}$ in
(\ref{CJ0S}) and (\ref{CJ1S}) with their analogues in earlier
quarkonium studies, namely the values of the squared radial 
quarkonium wave-functions at the origin, $|{\cal R}_{L}(0)|^2$, where
\begin{equation}
{\cal R}_{L}(0)\;\equiv\;\left.\frac{\partial^L}{\partial r^L} 
\,{\cal R}_{L}(r)\right|_{r=0}
\label{R1}
\;,
\end{equation}
with $r\equiv |{\bf r}|$.
Interrelating
$|{\cal R}_{L}(0)|^2$ and  ${\cal C}^2_{JLS}$ is somewhat ambigous,
because the latter coalescence probabilities are to be 
weighted with and summed over all possible $Q\overline{Q}$ configurations with
distinct relative momentum and spatial separation, while the former 
$|{\cal R}_{L}(0)|^2$ is just a number, independent of space-time
and the momentum distribution of $Q$ and $\overline{Q}$. Hence, one expects
the summed-over coalescence probabilities generally be larger.
 Nevertheless, one may establish as a limiting estimate an 
approximate correspondance between the ${\cal C}^2_{JLS}$ and the
$|{\cal R}_{L}(0)|^2$ by concentrating on the phase-space region
where coalescence receives its dominant contribution from,
namely at the maximum value. This is in accord with the commonly
used approximation of using the value of  is approximately given by
${\cal R}_{L}(r)$ at $r=0$ where the radial wave-function is maximal.

\begin{figure}
\epsfxsize=350pt
\centerline{ \epsfbox{p100.11} }
\vspace{0.5cm}
\caption{
         Typical shape of the coalesence probabilities  (\ref{CJ0S}) and (\ref{CJ1S})
         for the different angular momentum eigenstates $^{2S+1}L_J$ as a function of 
         $\overline{r}\equiv|{\bf \overline{r}}|$ for various values of 
         $\overline{k}\equiv|{\bf \overline{k}}|$. 
         The values of $\mu_L$ and $\alpha_L$ entering the expressions 
         (\ref{CJ0S}) and (\ref{CJ1S}) are determined from (\ref{munu}) an
         (\ref{param2}), using the RMS values of Table 1.
         \label{fig:fig6}
         }
\end{figure}
\bigskip

Consequently, one may write
\begin{equation}
{\cal C}^2_{JLS}
\left(\mu^2 \,|\,  \overline{k}^2, \overline{r}_{max}^2(L), t_c  \right)
\;\widehat{=} \;
\frac{3}{4\pi\;(M_Q)^{2L}}\;\,
\int_{R_{Q\overline{Q}}}d^3 r \;\left|{\cal R}_{L}(0)\right|^2
\;\left( 1 + O(\overline{k}^2)\right)
\label{R2}
\;,
\end{equation}
where the maximum values of the coalescence probabilities (\ref{CJ0S}) and
(\ref{CJ1S}) are
\begin{equation}
\overline{r}_{max}^2(L) \;=\;
\left\{
\begin{array}{c}
0 \;\;\;\,\mbox{for $L=0$} \\
\alpha_L \;\;\;\mbox{for $L=1$} 
\end{array}
\right.
\;,
\end{equation}
and $t=t_c$ is the corresponding time argument that can be set to zero,
since it is s dummy variable for the present execise.
The integration is over the typical bound-state size with radius
$R_{Q\overline{Q}}\equiv R_{Q\overline{Q}}(^{2S+1}L_J)$.
In the limit $\overline{k}^2 \rightarrow 0$ one may therefore
write for the cases $L=0$ and $L=1$, respectively,
\begin{equation}
{\cal C}^2_{J0S} \left(\mu^2 \,|\,  0, 0, 0) \right)
\;\stackrel{(k\rightarrow 0)}{\approx} \;
\left| {\cal R}_{0}(0)\right|^2
\; R_{Q\overline{Q}}^3
\;,\;\;\;\;\;\;\;\;\;\;\;\;\;\;\;\;
{\cal C}^2_{J1S} \left(\mu^2 \,|\,  0, \alpha_1, 0) \right)
\;\stackrel{(k\rightarrow 0)}{\approx} \;
\frac{M_Q^2}{\left|{\cal R}_{1}(0)\right|^2}
\; R_{Q\overline{Q}}^3
\;\;\;\mbox{for $L=2$}
\label{R3}
\;.
\end{equation}
Typical values for $| {\cal R}_{L}(0)|^2$ from QCD potential model
calculations \cite{Eichten95}, e.g. for Buchm\"uller-type wave-functions, 
are
\begin{equation}
\left| {\cal R}_{0}(0)\right|^2
\;=\; 
\left\{
\begin{array}{c}
0.81 \; GeV^3 \;\;\;\mbox{for $c\bar{c}$} \\
6.48 \; GeV^3 \;\;\;\mbox{for $b\bar{b}$}
\end{array}
\right.
\;\;\;\;\;\;\;\;\;\;\;\;\;\;\;\;
\frac{\left| {\cal R}_{1}(0)\right|^2}{M_Q^2}
\;=\;
\left\{
\begin{array}{c}
0.041 \; GeV^3 \;\;\;\mbox{for $c\bar{c}$} \\
0.057 \; GeV^3 \;\;\;\mbox{for $b\bar{b}$}
\end{array}
\right.
\label{R4}
\end{equation}
for $S$-wave and $P$-wave states, respectively, with $M_c=1.35$ GeV and $M_b=$5 GeV.
Comparing  these values with the coalescence probabilities in the limit
(\ref{R3}), one obtains, on account of the explicit expressions
(\ref{CJ0S}) and (\ref{CJ1S}),
and  using  the values of $R_{Q\overline{Q}}(^{2S+1}L_J)$ 
from Table 1, the following estimates for $S$-states and $P$-states, respectively:
\begin{equation}
\frac{1}{R_{Q\overline{Q}}^3}\;
\frac{
{\cal C}^2_{J0S} \left(\mu^2 \,|\,  0, 0, 0) \right)
}{
\left| {\cal R}_{0}(0)\right|^2
}
\;\simeq\; 
\left\{
\begin{array}{c}
1.23 \;\;\mbox{for $c\bar{c}$} \\
1.29 \;\;\mbox{for $b\bar{b}$}
\end{array}
\right.
\;\;,\;\;\;\;\;\;\;\;\;\;\;\;\;\;\;\;
\frac{M_Q^2}{R_{Q\overline{Q}}^3}\;
\frac{
{\cal C}^2_{J1S} \left(\mu^2 \,|\,  0, \alpha_1, 0) \right)
}{
\left| {\cal R}_{1}(0)\right|^2
}
\;\simeq\; 
\left\{
\begin{array}{c}
1.15 \;\;\mbox{for $c\bar{c}$} \\
1.30 \;\;\mbox{for $b\bar{b}$}
\end{array}
\right.
\label{R5}
\;.
\end{equation}
One sees that
these ratios for $S$-wave and $P$-wave states, respectively, indicate 
that the coalescence model used in this paper gives  around 15-30 $\%$
larger values for the formation probabilities of quarkonium than 
the widely used approximation based on the values of
the radial wave-functions at zero $Q\overline{Q}$ separation.
It has to be kept in mind, though, that the actual magnitude of the
coalescence probabilities can be significantly larger than in (\ref{R5})
when summed over all possible $Q\overline{Q}$ momentum
configurations and trajectories.
\bigskip
\bigskip
\bigskip

\subsection{Cross-sections}
\medskip

In order to make contact with observable quantities such as multiplicities,
relative abundances or spectra of produced quarkonium mesons,
one needs to integrate the space-time dependent total squred
amplitude (\ref{ansatz1}) over the complete event history of
a collision $A+B$, as well
as  over all unobserved particle states $X$.
For instance, the total production cross-section
for quarkonia $H_{Q\bar{Q}}(^{2S+1}L_J)$ is obtained after
dividing by th incoming parton flux 
and setting without loss of generality $t_0=0$, $\overline{{\bf R}}_0={\bf 0}$,
\begin{equation}
d\sigma\left(AB\rightarrow H_{Q\bar{Q}}+X\right)
\;=\;
\sum_{c^\ast=g,Q\bar{Q}}
\;
d\sigma\left(AB\rightarrow c^\ast \;X\right)(\hat{s},Q^2)
\;\times\;
\Omega_{c^\ast}^{JLS}\left(Q^2, \mu^2\right)
\;,
\label{dsig1}
\end{equation}
where
\begin{eqnarray}
d\sigma\left(AB\rightarrow c^\ast \;X\right)(\hat{s},Q^2)
&=&
\sum_{abd}
\,f_{a/A}(x_a,\mu^2)  \; f_{b/B}(x_b,\mu^2) 
\; S_a(x_a,\mu^2,Q^2 \,|\; 0)\, S_b(x_b,\mu^2,Q^2\,|\, 0)
\;T_d(Q^2,\mu^2\,|\,0)
\;\times
\nonumber \\
& & \nonumber \\
& &
\times\;
\left(\frac{}{}
\frac{1}{2 \hat{s}}
\,
\left| A(ab\rightarrow c^\ast d)\right|^2(\hat{s},Q^2)
\right)
\;\,
(2\pi)^4\delta(k_a+k_b-k_{c^\ast}-k_d) \;\,dx_a\,dx_b\,d^4k_{c^\ast} d^4k_d
\label{dsig2}
\;.
\end{eqnarray}
and
\begin{equation}
\Omega_{c^\ast}^{JLS}\left(Q^2, \mu^2\right)
\;=\; 
\int_{0}^{\infty}dt \;
\int d^3 \overline{r}_1 d^3 \overline{r}_2 \;
\int d^4 k_1 d^4 k_2 \;
{\cal E}^2_{c^\ast}\left(Q^2, \mu^2 
\,|\, k_1, k_2, \overline{{\bf r}}_1,\overline{{\bf r}}_2, t \right)
\;\;
{\cal C}^2_{JLS}
\left(\mu^2 \,|\, k_1, k_2, \overline{{\bf r}}_1,\overline{{\bf r}}_2, t \right)
\;.
\label{sig3}
\end{equation}
Using the kinematic variable definitions (\ref{def1})-(\ref{def4}),
one can express the triple-differential hadronic cross-section 
for $H_{Q\bar{Q}}$ production in terms of the rapidities
$y_1$ of the jet $c^\ast$ that initiates the $Q\bar{Q}$ evolution,
the rapidity $y_2$ of the recoiling jet $d$ which 
evolves and fragments by itself, and the common relative
transverse momentum $Q = p_\perp$ posessed by each of the two outgoing jets $c^\ast$ and $d$,
\begin{equation}
\frac{d\sigma}{dy_1dy_2dp_\perp^2}(AB\rightarrow H_{Q\bar{Q}}+X)
\;=\;
\sum_{c^\ast=g,Q\bar{Q}}
\;
\frac{d\sigma}{dy_1dy_2dp_\perp^2}(AB\rightarrow c^\ast \;X)
\;\times\;
\Omega_{c^\ast}^{JLS}\left(p_\perp^2, \mu^2\right)
\label{sig1}
\;,
\end{equation}
where $\Omega_{c^\ast}^{JLS}$ is the same as above,
and
\begin{eqnarray}
& &
\frac{d\sigma}{dy_1dy_2dp_\perp^2}(AB\rightarrow c^\ast \;X)
\;=\;
\sum_{abd}
\;x_a\,f_{a/A}(x_a,\mu^2)  \; x_b\,f_{b/B}(x_b,\mu^2) 
\;\;
\left(\frac{}{}
\frac{1}{16\pi \hat{s}^2}\; \left| A(ab\rightarrow c^\ast d)\right|^2(\hat{s},p_\perp^2)
\right)
\;\times
\nonumber \\
& & \nonumber \\
& &
\;\;\;\;\;\;\;\;\;\;\;\;\;\;\;
\;\;\;\;\;\;\;\;\;\;\;\;\;\;\;
\;\;\;\;\;\;\;\;\;\;\;\;\;\;\;
\;\;\;\;\;\;\;\;\;\;\;\;\;\;\;
\times\;
S_a(x_a,\mu^2,p_\perp^2\,|\, 0)\; S_b(x_b,\mu^2,p_\perp^2 \,|\; 0)\; T_d(p_\perp^2,\mu^2\,|\,0)
\;.
\label{sig2}
\end{eqnarray}
\bigskip
\bigskip

\subsection{Calculation scheme}
\medskip

Due to its probabilistic character, the above formalism is suitable for implementation 
in a Monte Carlo simulation using the computer program VNI \cite{ms44},
which allows one to
trace the microscopic history of the dynamically-evolving 
particle system
in space-time {\it and} momentum space, from the instant of  collision
of the beam particles $A$ and $B$ to the eventual emergence of quarkonium states
plus all other final-state particles.
Specifically, following the space-time development of the complete system
of particles in a collision event, embodies the dynamics of the $Q\overline{Q}$ 
system within the rest of the event, consisting of all other produced
partons and the hadronization products, as well as the fragmentation
of the spectator partons of the beam remants of the original $A$ and $B$.
This allows to account  not only
for the correlations of particles in space,  time, color and flavor, but
also for the statistical occurence of quarkonium formation by $Q\overline{Q}$
coalescence versus open charm/bottom production by recombination of
$Q$ and $\overline{Q}$ with partons of the rest of the event.

Referring for technical details to Ref. \cite{ms44}, I confine myself
here to a brief outline of the simulation scheme.
The time development in position and momentum space
of the space-time dependent admixture 
of partons and emerging hadronic excitations, including $Q$ and $\overline{Q}$
and its final hadronic products,
is obtained by following each individual particle through its 
history with
the various  probabilities and time scales of interactions 
sampled stochastically from the relevant cross-sections.
The detailed history of the system can thus be traced by evolving
the phase-space distributions of particles are
evolved in small time steps ($\Delta t \simeq 10^{-3}\;fm$)
 and 7-dimensional phase-space $d^3rd^3kdE$, from the instant of collision
between the beam particles $A$ and $B$, all the way 
until stable final-state hadrons and other particles
are left as freely-streaming particles.
At any time during the evolution, the state of the $Q\overline{Q}$ system
can be extracted, and so the final quarkonium yield at the end of of a
collisions event.
For the case of $pp$ and $p\bar{p}$ collisions on which I shall concentrate
in the following,
the essential ingredients in this Monte-Carlo procedure are summarized 
as follows \cite{ms44}:
\begin{description}
\item[(i)]
The {\it initial state} is constructed by
decomposing the incoming beam particles, $p$ and $\bar{p}$, into their
parton substructure according to measured structure functions
\footnote{
The GRV structure function parametrization \cite{grv},
is used throughout, which describes quite accurately the 
HERA data even at low $Q^2$ and very small $x$.
}
with a spatial distribution given by the Fourier transform 
of the proton elastic form-factor.
The so-initialized phase-space densities of
(off-shell) partons are then boosted with
the proper Lorentz factor to the center-of-mass frame of the
colliding beam particles.
\item[(ii)]
The {\it initial $Q\overline{Q}$ production} 
is obtained by triggering on
those hard parton collisions $ab \rightarrow c^\ast d$ that contain
a $c^\ast= Q\overline{Q}$ or  $c^\ast=g^\ast$ with $g^\ast \rightarrow Q\overline{Q}$.
These hard processes are sampled from the 
well-documented cross-sections \cite{CSM2,CSM3,Cho1}, including the direct process and
gluon fragmentation process with weights given by the relative magnitudes
of the corresponding cross-sections.
\item[(iii)]
The {\it parton cascade} development
proceeds then by
propagating all partons along classical trajectories until they interact
(radiate, coalesce, or hadronize),
including the $Q\overline{Q}$ system as well as te additionally produced
partons from initial-state space-like radiation off $a,b$ and final-state
time-like radiation off the $Q\overline{Q}$ system and the recoiling jet $d$.
The production of each parton through the hard scattering or  radiation 
is subject to an individual, kinematic-specific 
formation time
$\Delta t_{p,c} = \gamma / M_{p}$ where $1/M_{p}=1/\sqrt{k^2}$ is the 
proper decay time of an off-shell  parton with invariant mass $M_p$
and $\gamma=E/M_{p}$ is the Lorentz factor.
\item[(iv)]
The {\it final Q$\overline{Q}$ quarkonium formation} 
is traced by evaluating in each infinitesimal step of the
collision event the  spectrum of coalescence probabilities (\ref{CJ0S}) and
(\ref{CJ1S}) from the $Q$ and $Q\overline{Q}$ positions and momenta, 
and testing the values against a random number. At each point in time
the total probability determines whether a quarkonium state may be produced,
and if so, the relative probabilities of the different states give the
weights for which quarkonium state out of the spectrum is formed.
If no coalescence occurs, the $Q$ and $\overline{Q}$ continue in their
evolution as indiviual particles as in (iii).
\item[(v)]
The  {\it emergence of hadrons other than quarkonium} involves the recombination
of produced partons into pre-hadronic cluster, using the Ellis-Geiger scheme \cite{EG},
in which at any point  during the cascade dvelopment any pair of nearest-neigbor partons
is tested whether their mutual separation favors a cluster formation,
or whether they proceed in their parton evolution as in (iii).
These  pre-hadronic clusters decay subsequently into primary hadrons, which in turn
if excited or resonant states, can decay into final stable particles according
to the particle data tables \cite{pada}.
Again, each newly produced pre-hadronic cluster as well
as emerging hadron becomes a `real' particle only after
a characteristic formation time 
$\Delta t_{c,h} = \gamma / M_{c,h}$ depending on the invariant mass $M_{c,h}$ and
the energy through $\gamma= E/M_{c,h}$. Before that time has passed, a 
cluster or hadron
is considered as a still virtual object that cannot interact
incoherently until it has formed according to the uncertainty principle.
\item[(vi)]
The {\it beam remnants}, being the unscathed remainders of the initial beam particles
(in the present case $p$ and $\bar{p}$), 
emerge  from reassembling all those  remnant primary partons that
have been spectators without interactions throughout the evolution.
The recollection of those yields two corresponding beam clusters
with definite charge, baryon number, energy-momentum and position, as given by
the sum of their constituents.
These beam clusters  decay into final-state hadrons which recede
along the beam direction at large rapidities of the beam/target fragmentation 
regions.
Again, individual formation times of the produced hadrons are accounted for. 
\end{description}
It is important to realize that the spatial density and the momentum distribution
of the particles are intimately connected: The momentum 
distribution continuously changes through the interactions 
\footnote{
For the case of 
$pp$ or $p\bar{p}$ collisions considered in the following,
the general variety of particle interactions reduces to only the 
radiative parton poduction associated with the evolution before and
after the hard scattering, and the parton-hadron conversion through
parto-parton recombination or coalescence.}
and determines how the quanta propagate in coordinate space.
In turn, the probability for subsequent interactions depends on the 
resulting local particle density. Consequently, the development
of the phase-space densities is a complex
interplay, which - at a given point of time - contains implicitly the
complete preceding history of the system.
\bigskip
\bigskip
$\frac{}{}$\vspace{-3.0cm}
\begin{center}
\begin{minipage}[t]{15.0cm}
\begin{table}[t]
\epsfxsize=490pt
\centerline{ \epsfbox{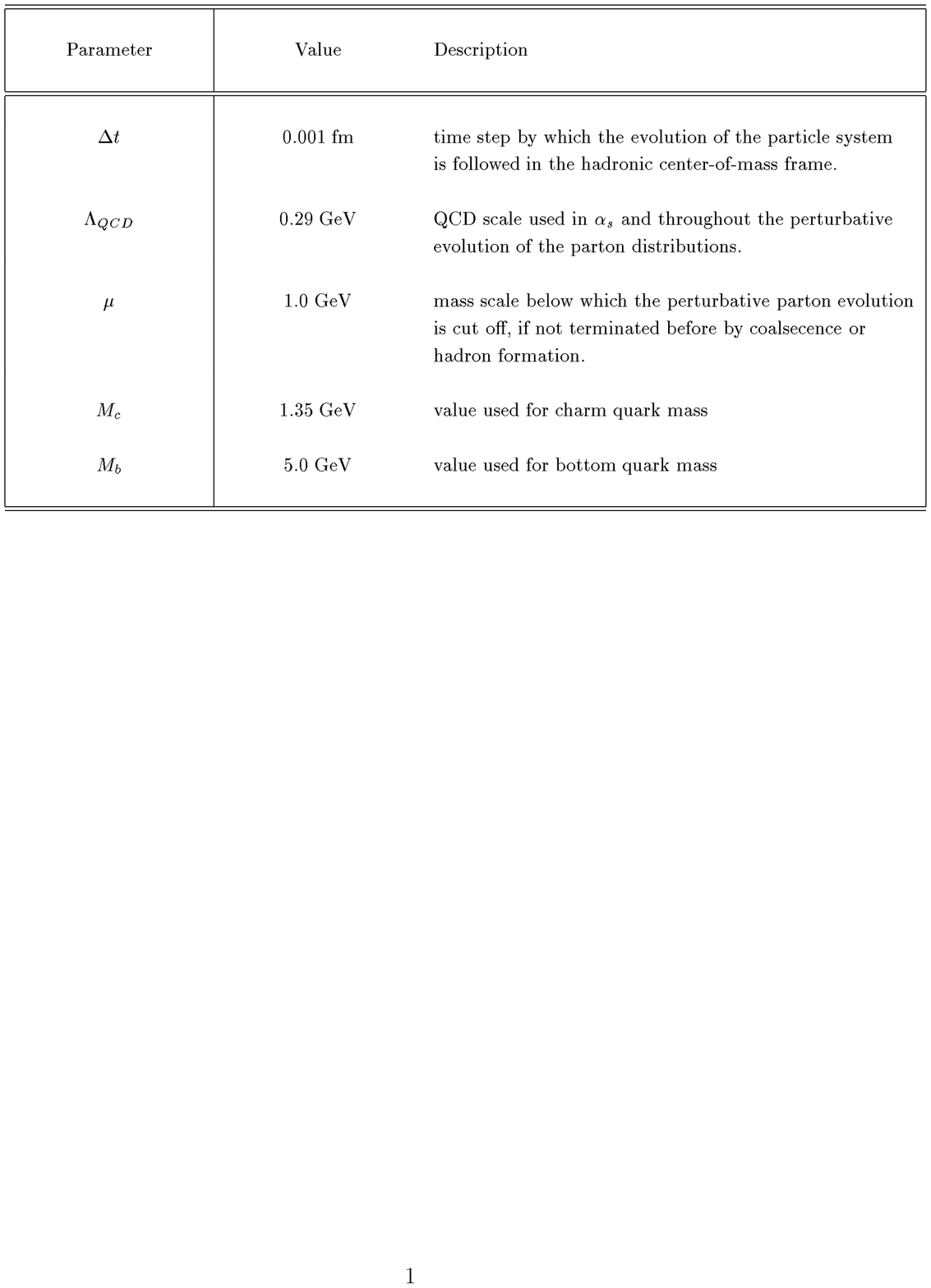} }
\vspace{-10.5cm}
\caption{Parameter values used in the model calculations.}
\end{table}
\end{minipage}
\end{center}
\newpage

\section{APPLICATION TO {\boldmath $pp (p\bar{p})$} COLLISIONS
AT {\boldmath $30\; GeV$ - $14\; TeV$}}
\label{sec:section3}
\bigskip

For the production of the charmonium and bottonium $S$-wave states 
$J/\psi$ and $\Upsilon$, respectively, there exists a body of
experimental data, both from $pp$ collisions at the ISR ($\sqrt{s}=$ 30-63 GeV)
and from $p\bar{p}$ experiments at the Tevatron ($\sqrt{s}=$ 1.8 TeV).
Also, the charmonium $P$-wave states $\chi_{c}$ have been observed, although with
larger experimental uncertainty, since these states are only indirectly measured
through their radiative decay into $J/\psi$ (unlike the $S$-wave states
which due to their narrow resonance character can be observed directly through 
invariant mass analysis).
On the other hand, 
for radial $S$-wave excitations, such as $\psi'$, $\Upsilon'$, $\Upsilon''$, 
data are more scarce, and detailed measurements of $\chi_b$ as well as 
of the $^1S_0$ and $^1P_1$ states, $\eta_c$, $\eta_b$, $h_c$, $h_b$,
are not reported at all.

Hence, I will  focus  primarily on the production of the charmonium states
$J/\psi$ ($^3S_1$) and $\chi_{c\,J}$ ($^3P_J$), as well as the
lowest-lying bottonium state $\Upsilon$ ($^3S_1$).
However, since in the model calculations the production patterns
of the various quarkonium states are intimately entwined by the
dynamical evolution, I also discuss somewhat the relevance of 
the other $S$-wave and $P$-wave states.
\bigskip
\bigskip

\subsection{Production cross-sections:  postdictions and predictions}
\medskip

The {\it principal result} of this paper is contained in Fig. 7, 
which collects  the  quarkonium production cross-sections for 
$J/\psi$, $\chi_c$ and $\Upsilon$, and which is
posted here forefront as basis for the subsequent analysis
of the underlying dynamics in the present approach.
Shown in Fig. 7 is the $p_\perp$-dependence of the cross-sections around central
rapidity at 63 GeV, 1.8 TeV and 14 TeV, as given by eq. (\ref{sig1}),
\begin{equation}
\left. \frac{d\sigma}{dp_\perp}\right|_{y=0}\;=\;
2 p_\perp\;\left. \frac{d\sigma}{dp_\perp^2}\right|_{y_1+y_2 \le 1}
\;\;\;\;\;\;\;\;\;\mbox{for}
\;\;\;pp (p\bar{p})\;\longrightarrow\;\;
\left\{
\begin{array}{l}
\; J/\psi \;+ \;X
\\
\; \chi_{c}\;\rightarrow J/\psi \;+ \;X
\\
\; \Upsilon \;+ \;X
\end{array}
\right.
\;,
\label{sig4}
\end{equation}
multiplied by the leptonic branching ratios $B(J/\psi\rightarrow \mu^+\mu^-) = 5.97\,\%$, 
and $B(\Upsilon\rightarrow \mu^+\mu^-) = 2.48\,\%$. 
Confronted are the calculated cross-sections (from eqs. (\ref{sig1}) and (\ref{sig2}))
with the measured spectra from Refs. \cite{tevatron,ISR}.
In order to allow direct comparison with
experimental data, the following quarkonium yields are plotted:
The top panel of Fig. 7 displays the production of  prompt $J/\psi$'s only,
with the radiative feed-down from $\chi_c$'s as well as from bottom decays 
subtracted. The middle part reflects the production of $\chi_c$ summed
over all $J=0,1,2$ states, and implicitely obtained through the amount of
$J/\psi$'s that result exclusively from the radiative decays of $\chi_c$
with the branching ratios
$B(\chi_{c\,0}\rightarrow J/\psi+\gamma) = 6.6\cdot 10^{-3}\,\%$, 
$B(\chi_{c\,1}\rightarrow J/\psi+\gamma) = 27.3\,\%$, 
$B(\chi_{c\,2}\rightarrow J/\psi+\gamma) = 13.5\,\%$.
Finally, the bottom panel shows the production of  $\Upsilon$'s,
including  the radiative feed-down from the $\chi_b$ states.
\smallskip

The features of the results in Fig. 7 may be itemized as follows:
\begin{description}
\item[(i)]
The agreement of the calculated cross-sections for
$J/\psi$, $\chi_c$ and $\Upsilon$ production with the experimental
data at ISR and Tevatron is very good.
Not only the $p_\perp$-shapes of the cross-sections,
but also the overall magnitudes are {\it post}dicted correctly for ISR and Tevatron.
The LHC curves may therefore be seen as quantitative {\it pre}dictions.
In view of the fact that the dynamics of  quarkonium production
in the model calculations is essentially
parameter-free (aside from the fixed quarkonium radii in Table 1),
this success suggests that the combination of  PQCD 
parton evolution with bound-state formation via coalescence
may in fact be a reasonable  visiualization of the underlying physics. 
\item[(ii)]
The overall pattern of charmonium and bottonium production is quite similar,
although with important differences in detail.
The comparison of the $J/\psi$ and $\Upsilon$ production reveals a
`flattening' of the $p_\perp$-spectra due to the larger quark mass in the latter
case (c.f. Table 2), and, connected with this,
a much smaller size of the $\Upsilon$ as compared to the $J/\psi$ (c.f. Table 1).
The confrontation of the $\chi_c$ and $J/\psi$ spectra implies a softer, i.e. `steeper',
$p_\perp$ tail in the $\chi_c$ production, which again is plausible due
to the larger size of the $\chi_c$ states.
The `flattening' indicates a power-behavior in the transverse
energy $\sqrt{4M_Q^2+p_\perp^2}$ rather than in the transverse momentum
$p_\perp$ itself.
\item[(iii)]
Most prominent in the charmonium sector is $\chi_c$ production, leading
also to a strong $\chi_c$ component in the inclusive $J/\psi$ yield through
radiative feed-down $\chi_c\rightarrow J/\psi+\gamma$. In fact, 
the calculations show that more than
30 $\%$ of the total $J/\psi$ yield originates from $\chi_c$ decays.
On the bottonium sector this pattern is repeated, 
as the model calculations indicate. Again, the $\chi_b$'s provide the largest
yield, however, they do not contribute as strongly to the $\Upsilon$ population
as in the corresponding case of $\chi_c$ and $J/\psi$. This is due to
to the larger quark mass which leads to a kinematic suppression at large $p_\perp$.
\end{description}
\newpage
$\frac{}{}$\vspace{0.5cm}
\begin{figure}
\epsfxsize=390pt
\centerline{ \epsfbox{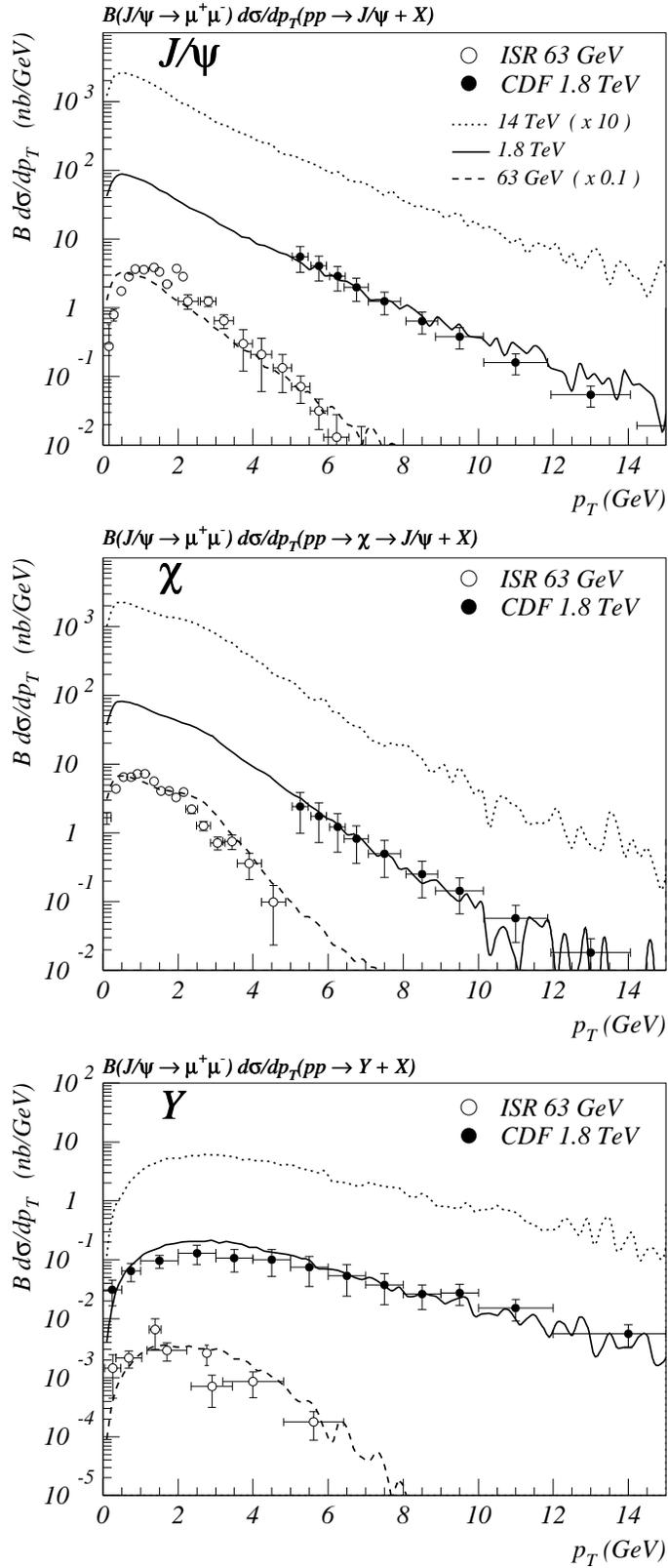} }
\vspace{1.5cm}
\caption{
         Comparison of calculated and measured  cross-sections for
         quarkonium production around central
         rapidity at 63 GeV, 1.8 TeV and 14 TeV. Top part includes
         prompt $J/\psi$'s only, with the radiative feed-down from $\chi_c$'s 
         as well as from bottom decays subtracted. Middle part shows  $\chi_c$ 
         production summed over all $J=0,1,2$ states, identified from
         $J/\psi$'s that result exclusively from the $\chi_c$ decays.
         Bottom part contains  $\Upsilon$'s, including  the radiative feed-down 
         from the $\chi_b$ states.
         For better distinction, the cross-sections have been scaled by
         factors of 0.1 (63 GeV), 1 (1.8 TeV) and 10 (14 TeV).
         The full curves are the model results. The data points are from
         Ref. [34] for ISR at 63 GeV and from Ref. [1] for Tevatron at 1.8 TeV.
         \label{fig:fig7}
         }
\end{figure}
\newpage

\noindent
In view of the non-trivial physics of quarkonium
production as discussed in Sec. I,
the good agreement of the model calculations with the available data
immediately poses the the question, 
\medskip

\noindent
\parbox{3.0cm}{}
\ $\;\;\;\;\;\;$
\parbox{16.0cm}{
{\it
Why does this space-time description, which  combines PQCD parton
production and evolution with a  simple coalescence model, 
agree  so well with experiment?
}
}
\medskip

\noindent
More specifically, there are three main issues to be examined in the following,
namely,
\begin{itemize}
\item
{\sl Initial $Q\overline{Q}$ production}:
What is the underlying production pattern, in particular, the relative
importance of color-singlet and color-octet contributions?  
\item
{\sl Parton evolution}:
How does the PQCD  evolution and the associated gluon radiation
affect the quarkonium yield?
\item
{\sl Coalescence}:
How does the space-time development of the initial $Q\overline{Q}$
proceed and what are
the characteristics of the final $Q\overline{Q}$ upon coalescence?
\end{itemize}
\bigskip
\bigskip

\begin{figure}
\epsfxsize=320pt
\centerline{ \epsfbox{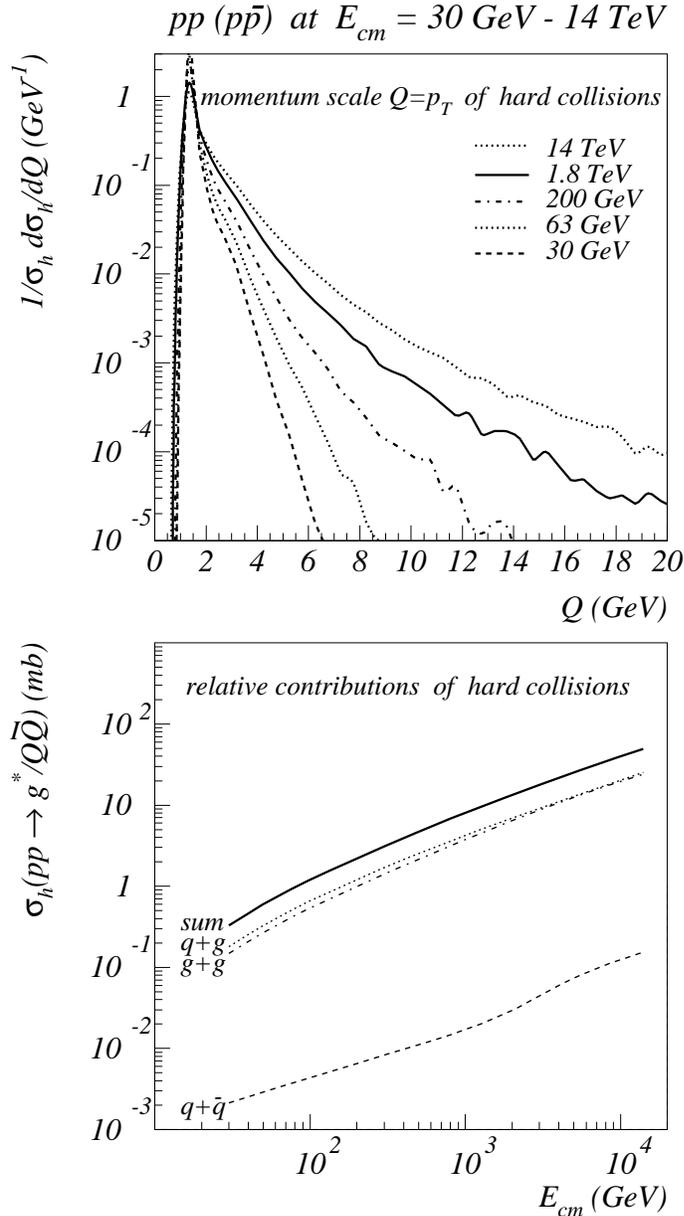} }
\vspace{-0.0cm}
\caption{
         {\sl Top panel}:
         Distribution of $Q^2$ values for the $pp$ $(p\bar{p})$   energy range  
         $\sqrt{s}=$ 30 GeV - 14 TeV, where  $Q \equiv p_\perp$ represents  the  
         momentum transfer at the hard vertex $ab \rightarrow c^\ast d$ of Fig 2. 
         {\sl Bottom panel}: 
         Integrated partonic cross-section for the production of the initial 
         $Q\overline{Q}$ from the  hard vertex $ab\rightarrow c^\ast d$ (directly, 
         or via gluon fragmentation) in th energy range  $\sqrt{s}=$ 30 GeV - 14 TeV.
         \label{fig:fig8}
         }
\end{figure}
\newpage

In the following Subsections these questions are addressed and discussed in
detail.
Beforehand, I would like to draw attention to 
Fig. 8, which may  serve 
to further some intuition for the beam-energy ($\sqrt{s}$) dependence of 
quarkonium production and the relative importance of
contributing partonic production processes.
The top part of Fig. 8 displays the distribution of $Q^2 \equiv p_\perp^2$
values for the $pp$ $(p\bar{p})$   beam energies 30 GeV - 14 TeV. It reflects the  
typical momentum transfers at the hard vertex $ab \rightarrow c^\ast d$, where
$a,b,d = g, q, \bar{q}$ and 
$c^\ast =  Q\overline{Q}$ (Fig. 2a) or
$c^\ast = g$ with $g\rightarrow  Q\overline{Q}$ (Fig. 2b).
As expected, there is a substantial increase in the mean $Q^2$, accompanied
by an increasingly prominent power-law tail, as one moves from ISR energy
30 GeV up to LHC energy 14 TeV. The immediate consequences are,
firstly, an enlarging of the kinematic region for producing heavier quarkonium states,
and secondly, the increasing importance of gluon radiation, the role
of which I shall discuss in detail later.
The bottom part of Fig. 8 shows the beam-energy dependence of
the integrated partonic cross-section for the production
of the initial $Q\overline{Q}$ from the  hard vertex $ab\rightarrow c^\ast d$
(directly, or via gluon fragmentation).
Three  features are self-evident here. First, the cross-section
rises by almost two orders of magnitude in the range 30 GeV - 14 TeV.
Second, the channel $ab=q\bar{q}$ of light $q\bar{q}$ annihilation
is down by 2 orders of magnitude, and is therefore completly irrelevant
in the energy range considered here.
Third, the contributions of $ab = qg$ and $ab = gg$ are about equal in
magnitude throughout, with the $qg$-channel slightly dominating.
This is a very important property that is often overlooked, because
one would naively expect that the $gg$-channel dominates by far due to
the strong presence of low-$x$ gluons in the protons.
However, collsions of low-$x$ gluons are suppressed in the $Q\overline{Q}$
production, because of the failure to overcome the mass threshold.
\bigskip
\bigskip

\subsection{Properties of color-singlet versus color-octet contributons}
\medskip

In addressing the first of the above issues, namely the relative importance of 
color-singlet versus color-octet contributions to the final-state quarkonium
yield, recall from the Sec. I, that historically
the `color-singlet model' \cite{CSM1,CSM2,CSM3}  did very well at ISR energies and below, but 
underpredicts badly the quarkonium productions at the Tevatron by 
more than an order of magnitude,
and thereby suggesting a strong `color-octet mechanism' \cite{RevBraaten,Cho1,Cho2} contributing
at very high energies.

In the present approach, both color-singlet and color-octet production
of the initial $Q\overline{Q}$ are self-consistently included, 
by determining  the
relative probabilities for singlet and octet production
from the respective weights of the hard $Q\overline{Q}$-production cross-sections.
It is important to stress again that an initial singlet (octet)
$Q\overline{Q}$ pair produced by the hard parton collision 
does not necessarily result in a final $Q\overline{Q}$ with
the same quantum numbers upon coalescence, because the emission
of gluons during the evolution changes the color state (and in principle
also spin and angular momentum) of the pair. In fact, since in the present
approach, the final $Q\overline{Q}$ has to be in a color-singlet state
in order to coalesce to a quarkonium bound-state, this is a crucial necessity
to be able to incorporate both singlet and
octet production of initial $Q\overline{Q}$ pairs.
Figs. 9 - 11 
separate the color-singlet and color-octet contributions and discriminate their
characteristic features.
\smallskip

Fig. 9 compares the relative proportions of the various included charmonium 
states at Tevatron energy $\sqrt{s} = 1.8$ TeV (the pattern is very similar on 
the bottonium sector, not shown here).
The top panel contains the realistic case with proper admixtures of both
color-singlet and -octet contributions, and shows a strong $J/\psi$ component,
yet being dominated by $\chi_c$ productions with a ratio $P(J/\psi)/P(\chi_c)\approx 2$.
The middle (bottom) panel exhibits the contributions 
from initial $Q\overline{Q}$  color-singlet (color-octet) pairs alone.
The patterns are strikingly different, as is already visible in the
earlier Fig. 4.
Color-singlet production of the initial $Q\overline{Q}$ results to
largest extent in $\chi_c$-formation (mostly $\chi_{c\,1}$),
whereas $J/\psi$, $\eta_c$ and $h_c$ are down by  far more than an
order of magnitude.
Contrary to that, when the initial $Q\overline{Q}$ is produced in a
color-octet, the situation is almost reversed. $J/\psi$ production
is the largest among the individual states, and is about equal to the sum of
all three $\chi_c$ states.
\smallskip

Figs. 10 and 11 detail the beam energy ($\sqrt{s}$) dependence
of color-singlet and -octet contributions for the charmonium and bottonium
sector, respectively.
In Fig. 10 the relative probabilities for $\chi_c$ and $J/\psi$
production are shown in the top panel, whereas the actual cross-sections
are plotted in the bottom panel $-$ both against the $pp$ ($p\bar{p}$)
center-of-mass energy $E_{cm} = \sqrt{s}$.
Correspondingly, Fig. 11 depicts the behavior of $\chi_b$ versus $\Upsilon$
production.
The fat lines are the sum of color-singlet and color-octet
contributions, the thin dashed (dotted) lines represent the color-singlet (-octet)
contributions alone.
From Fig. 10 one reads off that, in the present model,
the color-singlet component of $J/\psi$ is almost constant with $E_{cm}$, whereas  the
color-octet contribution rises strongly with increasing $E_{cm}$.
At the same time color-singlet and -octet contributions to $\chi_c$
are of the same order of magnitude, and decrease with $E_{cm}$.
As a result, at ISR energies 30 - 63 GeV,
the $J/\psi$ production is highly suppressed as compared to the $\chi_c$ yield,
at Tevatron energy 1.8 TeV, the two channels become almost comparable,
and at LHC energy 14 TeV, they become about equal.
A similar pattern is seen in Fig. 11 for $\chi_b$ and $\Upsilon$,
however with an even more pronounced energy dependence. Again, $\chi_b$
production is  by far favored  over $\Upsilon$ formation at lower energies,
but now already at Tevatron energy the two states are equally populated, and
at LHC energy the $\Upsilon$ production even dominates by a factor of $\simeq 1.5$.
\bigskip
\bigskip
\begin{figure}
\epsfxsize=450pt
\centerline{ \epsfbox{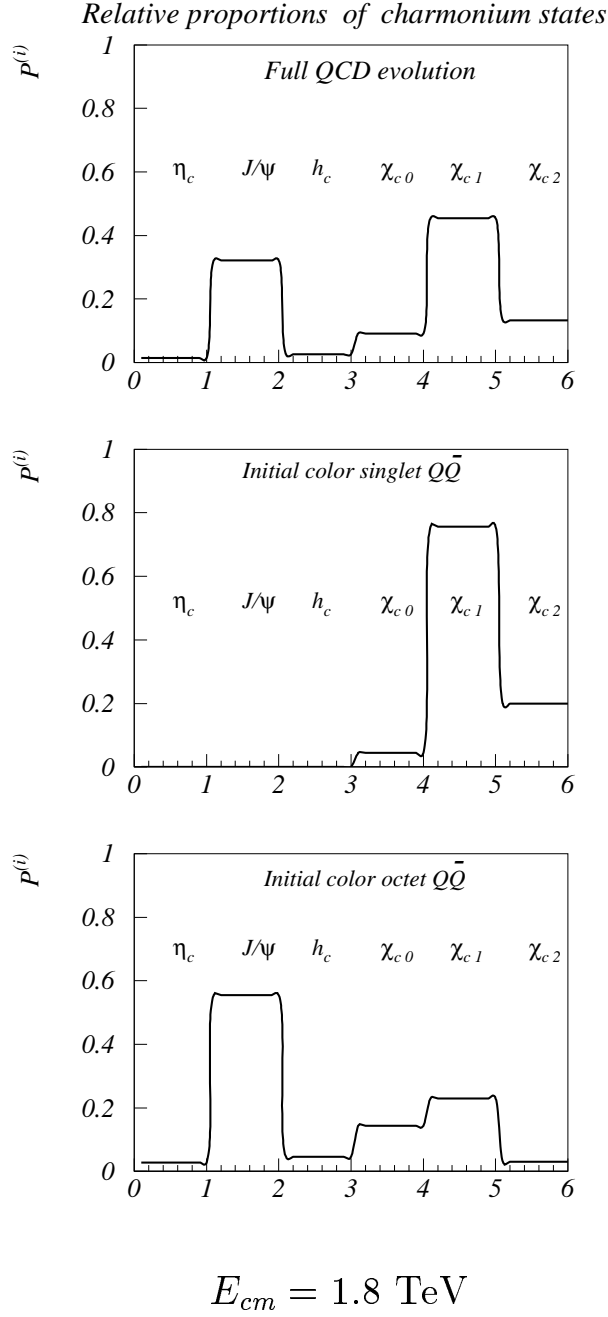} }
\vspace{-0.0cm}
\caption{
         Relative proportions of the various included charmonium 
         states at Tevatron energy $\sqrt{s} = 1.8$ TeV:
         {\sl top panel} contains the realistic case with proper admixtures of both
         color-singlet and color-octet contributions, 
         {\sl  middle panel} exhibits the contributions 
         from initial $Q\overline{Q}$  color-singlet pairs alone, whereas
         {\sl  bottom panel} displays the contributions 
         from initial $Q\overline{Q}$  color-octet pairs only.
         \label{fig:fig9}
         }
\end{figure}
\newpage
$\frac{}{}$\vspace{2.0cm}
\begin{figure}
\epsfxsize=350pt
\centerline{ \epsfbox{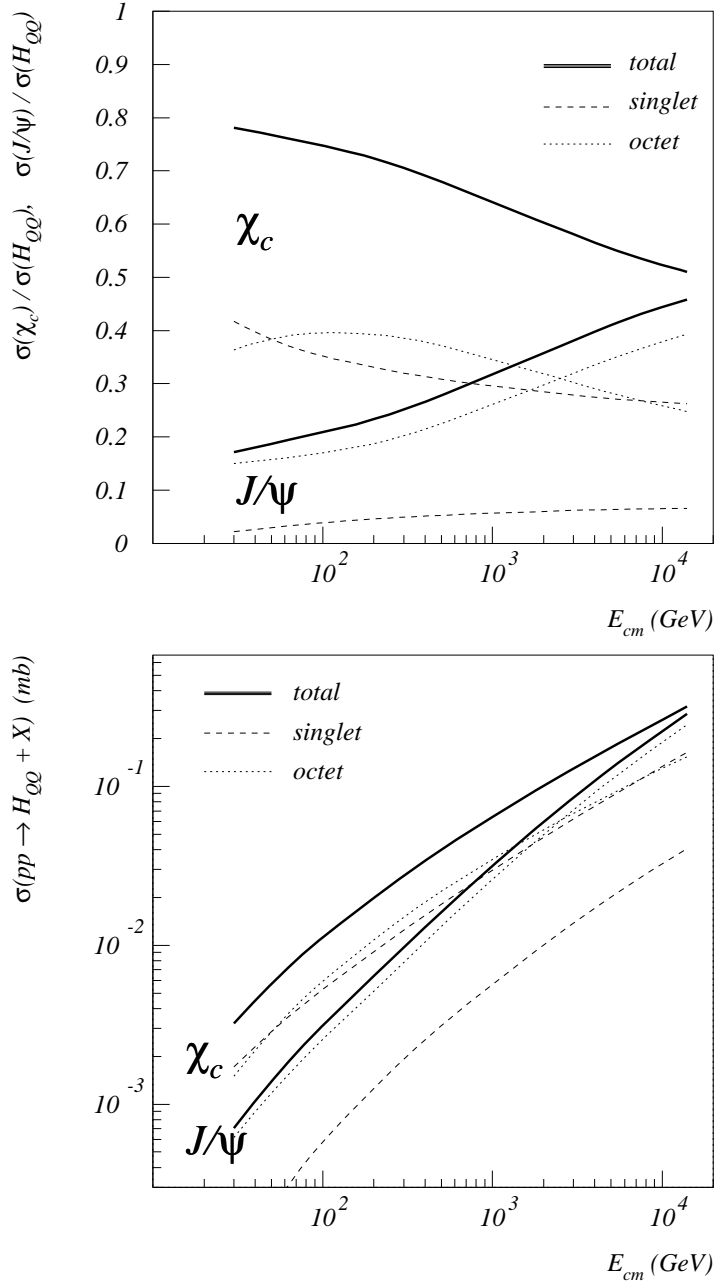} }
\vspace{0.5cm}
\caption{
         Relative probabilities for $\chi_c$ and $J/\psi$
         production (top), and corresponding  cross-sections
         (bottom) versus  $pp$ ($p\bar{p}$) center-of-mass energy $E_{cm} = \sqrt{s}$.
         The fat lines are the sum of color-singlet and color-octet
         contributions, the thin dashed (dotted) lines represent the color-singlet 
         (color-octet) contributions alone.
         \label{fig:fig10}
         }
\end{figure}
\newpage
$\frac{}{}$\vspace{2.0cm}
\begin{figure}
\epsfxsize=350pt
\centerline{ \epsfbox{p265.00} }
\vspace{0.5cm}
\caption{
         Relative probabilities for $\chi_b$ and $\Upsilon$
         production (top), and corresponding  cross-sections
         (bottom) versus  $pp$ ($p\bar{p}$) center-of-mass energy $E_{cm} = \sqrt{s}$.
         The fat lines are the sum of color-singlet and color-octet
         contributions, the thin dashed (dotted) lines represent the color-singlet 
         (color-octet) contributions alone.
         \label{fig:fig11}
         }
\end{figure}
\newpage

\subsection{Importance of parton evolution and gluon radiation}
\medskip

Next, I turn to the second question raised above which concerns the importance 
and effects of gluon radiation during the PQCD evolution.
As illustrated in Fig.2, gluon bremsstrahlung from both the virtual gluon 
(in case of the gluon fragmentation process)
and from the produced $Q\overline{Q}$ pair are included within the standard DGLAP evolution
plus angular-ordering in the leading log approximation \cite{DGLAP2}.
This (multi-)gluon emission is a most important aspect in the present approach.
Firstly, it changes the color-state of the $Q\overline{Q}$ which makes it possible
for an initial $Q\overline{Q}$ color-octet to evolve into a final
$Q\overline{Q}$ color-singlet and to coalesce to a physical bound-state.
\footnote{
Notice that a color-octet $Q\overline{Q}$ produced in a $^3S_1$ state
must radiate at least two gluons in order to convert into a color-singlet
state with quantum numbers of $J/\psi$ or $\Upsilon$. On the other hand,
if it is produced in a $3^P_J$ state, the emission of only one
gluon is necssary to form a $\chi_{c\,J}$ or $\chi_{b\,J}$.
}
Without gluon emission, only those $Q\overline{Q}$ pairs that are initially produced
as color-singlets could contribute to quarkonium formation. 
Secondly, gluon radiation causes significant energy loss of the $Q\overline{Q}$
system at high energy,
causing the relative momentum of the pair to be smaller in the mean, and their
relative distance to increase slower. In effect, this dynamical feature enhances the
probability for coalescence, since configurations with very small relative
momentum and separation are strongly favored (c.f. Fig. 6). 
In other words, the presence of gluon radiation is crucial, because
it allows the natural inclusion of both color-singlet and color-octet 
initial $Q\overline{Q}$ pairs to evolve into quarkonium,
and because the coalescence of final $Q\overline{Q}$ systems is more likely
to occur after sufficient gluon radiation to reduce the relative velocity to a value
commensurate with the non-relativistic kinematics of the accessible 
quarkonium bound-states.
\smallskip

Fig. 12 clearly displays the effect of gluon radiation in the model calculations.
Confronted are the production cross-sections (\ref{sig4}) for the
realistic scenario including gluon emission during the PQCD evolution 
which are the same as in Fig. 7 (fat lines),
with the corresponding cross-sections obtained by switching off
all gluon radiation (thin lines).
The comparison of the calculations with and without multiple
gluon emission reveals two substantial differences.
First, the overall magnitude in the case without gluon radiation
is significantly lower by a factor of about  3  for $J/\psi$ and $\Upsilon$,
but only slightly less for $\chi_c$.
Second, the shapes of the $S$-wave states $J/\psi$ and $\Upsilon$ are quite
different in the two scenarios, indicating a softer, i.e. steeper,
$p_\perp$-dependence in the no-radiation case. The shape of 
the cross-section for the $P$-wave $\chi_c$, on the other hand,
shows a suppression at low $p_\perp$ in the case without
gluon radiation, but is hardly affected at large $p_\perp$.

Both these points can be intuitively understood.
The substantial under-population of $J/\psi$ and $\Upsilon$ in the no-radiation
case is due to the fact that without gluon emision, only the color-singlet component
of initial $Q\overline{Q}$ production contributes, because
color-octet pairs have no chance to convert into a final-state
color-singlet, as is necessary to form quarkonium.
Since at $\sqrt{s}=$ 1.8 TeV the color-singlet cross-sections
are smaller than the color-octet counterparts (c.f Figs. 10 and 11),
the yield of $J/\psi$ and $\Upsilon$ is naturally underestimated.
On the other hand, the $\chi_c$ production is much less deviating in the
two cases, because here color-singlet and color-octet cross-sections are roughly
of equal magnitude.

The main lesson  learned from Fig. 12 is that the neglect of
gluon radiation results in 
quarkonium cross-sections which are incompatible with the data shown
earlier in Fig. 7, both with respect to the absolute yield
and the shape of the $p_\perp$-dependence.
\bigskip

Fig. 13 further confirms the importance of multi-gluon
emission. Plotted are the probability  distributions
of the number of  gluons being emitted 
during the PQCD evolution between the hard vertex $ab\rightarrow c^\ast d$
of Fig. 2, up to the point of coalescence to a quarkonium bound-state.
The full (dashed) histograms are for charmonium (bottonium) at
$pp$ ($p\bar{p}$ center-of-mass energies
63 GeV (top), 1.8 TeV (middle), and 14 TeV (bottom).
One observes that  $b\bar{b}$ evolution is generally accompanied by 
a larger amount of gluon radiation than $c\bar{c}$, an effect that increases
with energy.
Even at ISR energy 63 GeV, gluon radiation is non-negligible,
with mean values of $\langle N_g \rangle = 1.4$ (1.7) for
$c\bar{c}$  ($b\bar{b}$).
At Tevatron energy 1.8 TeV, the mean number of gluons emitted per $Q\overline{Q}$
system are
$\langle N_g \rangle = 1.9$ (4.7) for $c\bar{c}$  ($b\bar{b}$),
and at LHC energy 14 TeV, one has
$\langle N_g \rangle = 2.1$ (5.8).
This shows that even at ISR energy 63 GeV, gluon radiation is non-negligible,
and that it is most essential at Tevatron and LHC.
\newpage
$\frac{}{}$\vspace{0.5cm}
\begin{figure}
\epsfxsize=390pt
\centerline{ \epsfbox{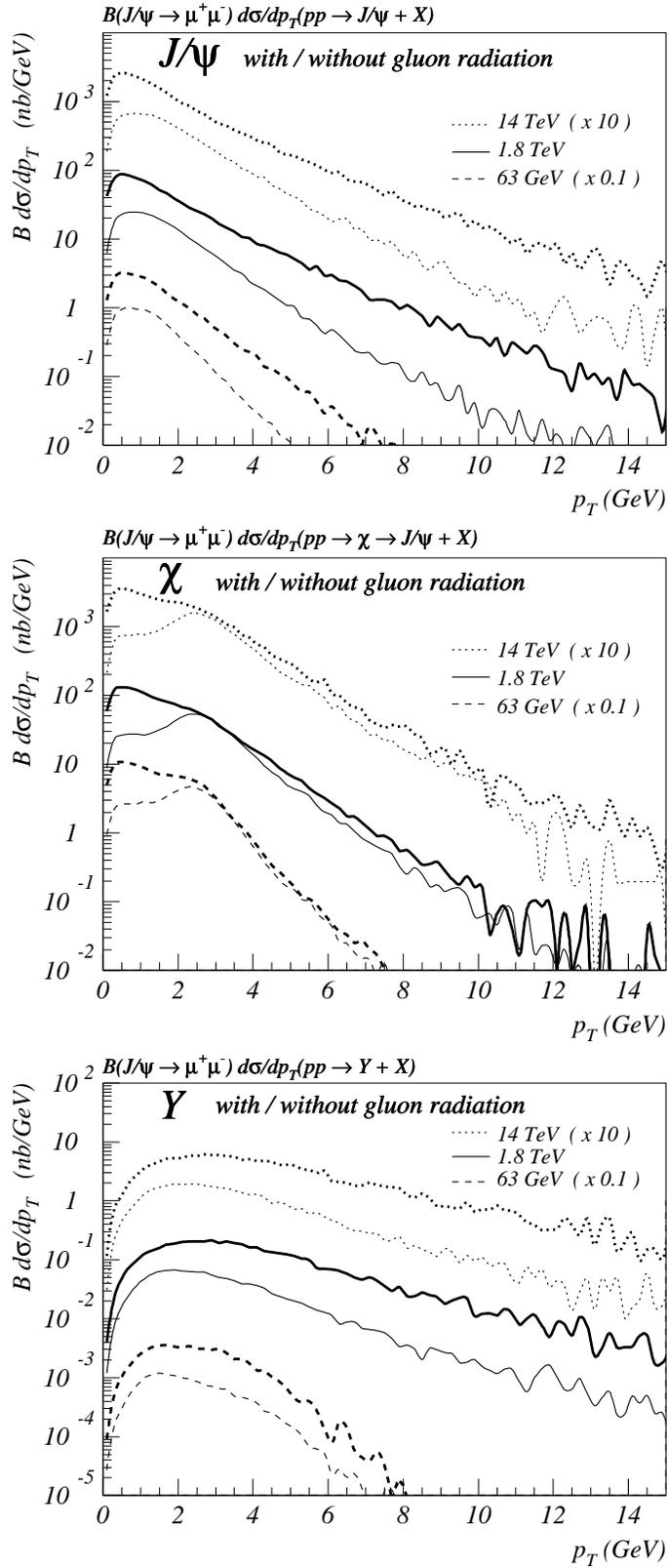} }
\vspace{1.5cm}
\caption{
         Quarkonium production cross-sections of the previous Fig. 7 revisited:
         Fat lines represent the  realistic scenario including gluon emission 
         during the PQCD evolution, whereas thin lines correspond to
         the no-radiation scenario with all gluon radiation switched off.
         Top part includes prompt $J/\psi$'s only,
         middle part shows  $\chi_c$'s  from  radiative $\chi_c\rightarrow \gamma+J/\psi$,
         and bottom part contains  $\Upsilon$'s, including  the radiative feed-down 
         from the $\chi_b$ states.
         \label{fig:fig12}
         }
\end{figure}
\newpage
$\frac{}{}$\vspace{1.0cm}
\begin{figure}
\epsfxsize=400pt
\centerline{ \epsfbox{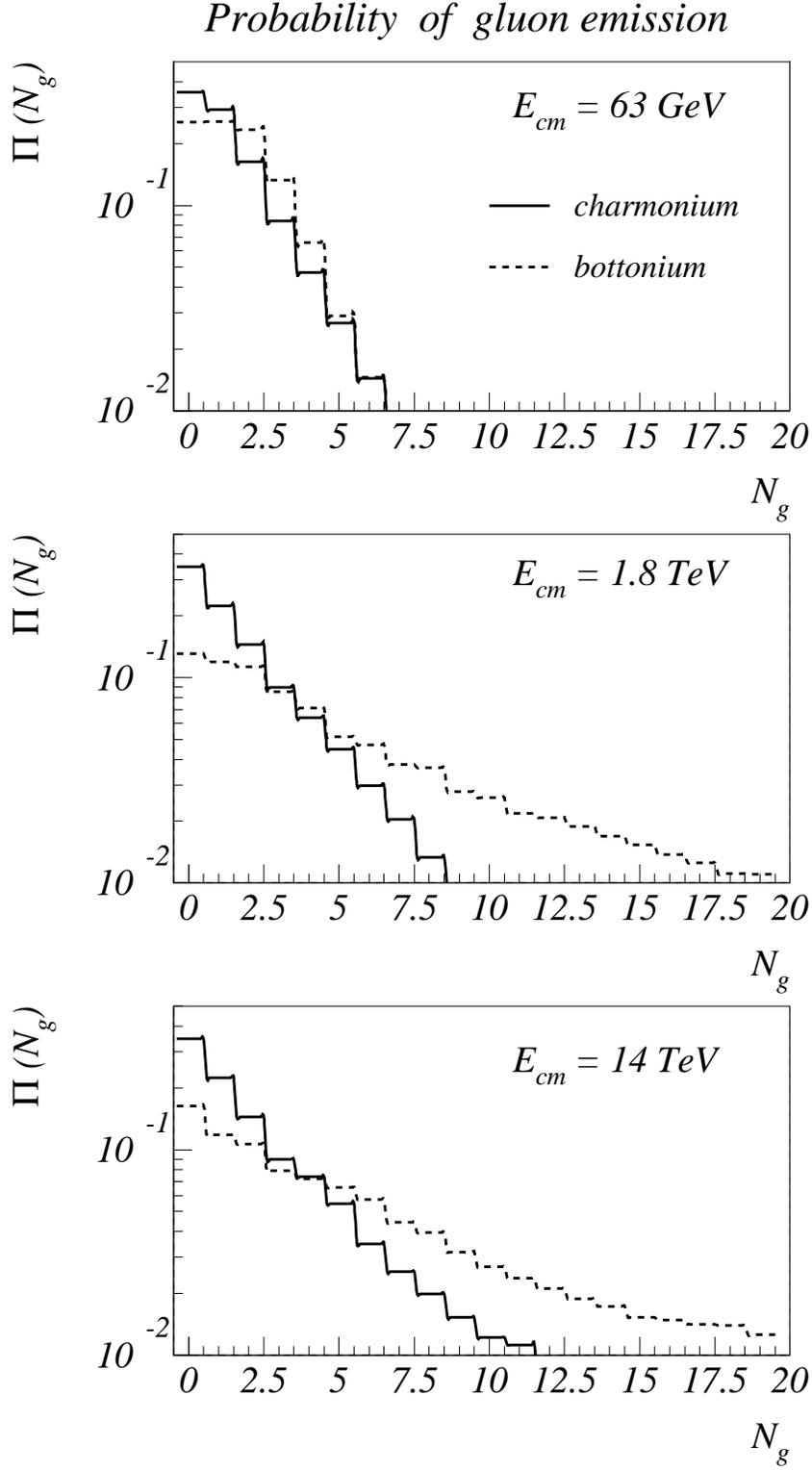} }
\vspace{0.5cm}
\caption{
         Probability  distributions of the average number of  gluons 
         being emitted during the PQCD evolution between the hard vertex 
         $ab\rightarrow c^\ast d$ of Fig. 2, up to the point of coalescence 
         to a quarkonium bound-state.  The full (dashed) histograms are for 
         charmonium (bottonium) at $pp$ ($p\bar{p}$ center-of-mass energies
         63 GeV (top), 1.8 TeV (middle), and 14 TeV (bottom).
         \label{fig:fig13}
         }
\end{figure}
\newpage

\subsection{Characteristics of the space-time development}
\medskip

I finally address the third and last question raised at the end of Sec. III.A,
namely, how does the time evolution of the $Q\overline{Q}$ system proceed
after being initially produced either directly at the hard vertex (Fig. a),
or by virtual gluon decay (Fig. 2b), as it propagates in space up to the point where it either
forms a bound-state, or ends up in other hadrons as open charm, respectively bottom.
In view of the preceding discussions, I wish to detail some characteristic 
space-time features that are related to the issues of multi-gluon
emission and color-singlet/color-octet processes.
\bigskip
\bigskip
$\frac{}{}$\vspace{1.0cm}
\begin{figure}
\epsfxsize=300pt
\centerline{ \epsfbox{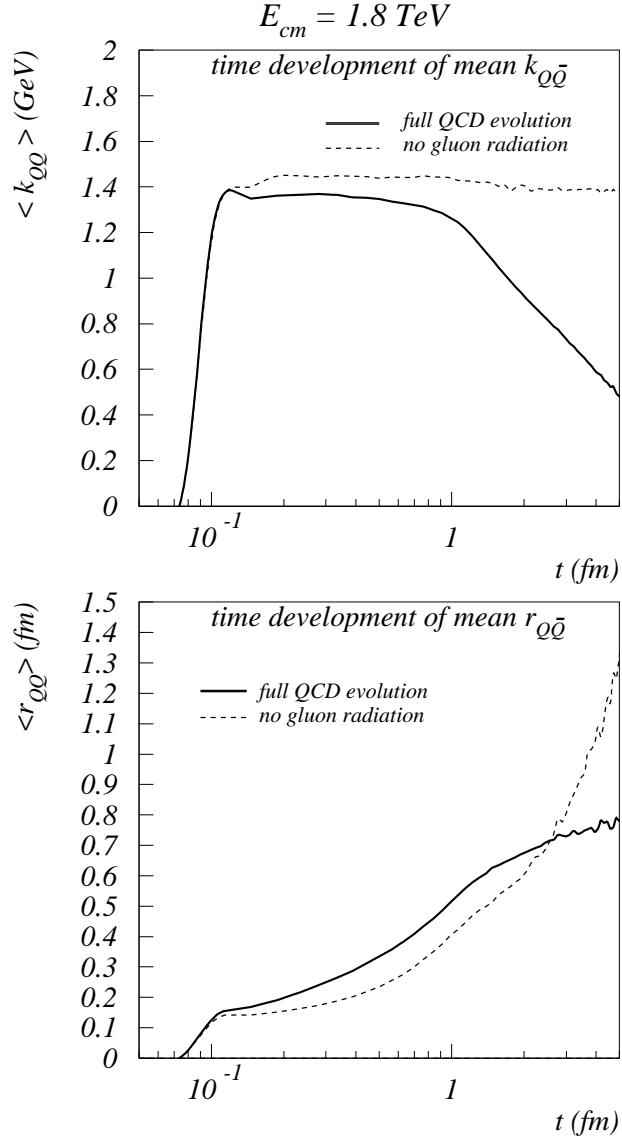} }
\vspace{0.5cm}
\caption{
         Time-evolution of the average relative $Q\overline{Q}$-momentum
         $k_{Q\overline{Q}}\equiv \langle |{\bf k}_Q - {\bf k}_{\overline{Q}} | \rangle$
         (top) and of the average $Q\bar{Q}$-separation
          $r_{Q\overline{Q}}\equiv \langle |{\bf r}_Q - {\bf r}_{\overline{Q}} | \rangle$
         (bottom).
         Solid curves represent the `full QCD evolution' including gluon radiation,
         and dashed curves correspond to  `free propagation' without gluon radiation.
         The plots refer to charmonium production at  1.8 TeV and include
         only those $Q\overline{Q}$ that actually end up as final charmionium states.
         \label{fig:fig14}
         }
\end{figure}
\newpage

Fig. 14 vividly illustrates the different time-evolution patterns for
the `full QCD evolution' including gluon radiation (solid curves)
and the `free propagation' without gluon radiation (dashed curves),
here exemplarily depicted for charmonium production at Tevatron energy 1.8 TeV.
Only those $Q\overline{Q}$ are included in the plots that actually
end up as final charmonium states, i.e., open charm production has been subtracted.
The top panel shows the change of the average relative momentum
$k_{Q\overline{Q}}\equiv \langle |{\bf k}_Q - {\bf k}_{\overline{Q}} | \rangle$
of $Q$ and $\overline{Q}$, 
as a function of time in the hadronic center-of-mass frame.
The bottom panel displays the corresponding time-dependence
of the relative separation of $Q$ and $\overline{Q}$ in the pair rest-frame,
$r_{Q\overline{Q}}\equiv \langle |{\bf r}_Q - {\bf r}_{\overline{Q}} | \rangle$.
The two plots confirm what was mentioned already ealier,
namely that multiple gluon emission leads to a 
significant energy loss 
which results in a drop 
\footnote{
It should be noted that the pronounced curvature of
$k_{Q\overline{Q}}$ around 1 $fm$ arises as a consequence of the
formation time of emitted gluons, $t_{form}\propto E_g/k^2_{\perp\,g}$.
Since the bulk of bremsstrahlung is associated with
low-energy gluons, the cumulate effect of these latter on $k_{Q\overline{Q}}$ 
becomes significant only after the typical formation time
$ \, \lower3pt\hbox{$\buildrel >\over\sim$}\, 1$ $fm$.
}
of the relative momentum $k_{Q\overline{Q}}$
(transverse momentum broadening is comparably ineffective).
The natural consequence is a slower time change in the relative
pair distance $r_{Q\overline{Q}}$ in the case with gluon radiation.
Without the emission of gluons, the $Q$ and $\overline{Q}$ do not
alter their initially obtained momentum and propagate classically in free space,
which leads to a faster growth in their spatial relative  separation.

This behavior has interesting consequences for the coalescence process
and the resulting quarkonium yield.
As is evident from the previous Fig. 6, $Q\overline{Q}$ configurations
with small $r_{Q\overline{Q}}$ and small $k_{Q\overline{Q}}$ have the
largest coalescence probabilities, and hence are strongly preferred in the
formation of quarkonium bound-states.
This in turn explains why in Fig. 12 the no-radiation scenario gives
a lower quarkonium cross-section than the case with gluon radiation.
\medskip

Fig. 15 exhibits, again for the charmonium sector at 1.8 TeV,
further striking differences between the `full QCD evolution' including gluon radiation,
and the `free propagation' without gluon emission.
The top part shows the time rate of change of the probability for
$Q\overline{Q}$ bound-state formation, versus the probability for $Q$ and $\overline{Q}$
to disappear individually in open charm (i.e., in other  charmed hadrons by
recombination with partons from the rest of the $p\bar{p}$ event).
The bottom part depicts for those $Q\overline{Q}$ pairs that indeed
form bound-states, the time-dependence of the individual
coalescence probabilities $\langle \Pi^{(i)}(t) \rangle$ for the
spectrum of charmonium states  $i = ^1S_0, ^3S_1, ^1P_1, ^3P_J$.
Both plots exhibit  very different shapes for the two cases, with and without
gluon radiation.
Although this time-development is not directly accessible in experiments,
one may think of sensible observables that reflect some of these features
and would allow to study the space-time structure by indirect measurement.
\medskip

Finally, Fig. 16 gives an instructive summary of the space-time history
that the initially produced $Q\overline{Q}$ pairs have experienced
on their way to coalescence as final $Q\overline{Q}$ color-singlet systems.
Again, I choose as example the charmonium sector at  $p\bar{p}$
center-of-mass energy of 1.8 TeV.
The left side shows the distributions in relative separation
$r_{ini}$, relative momentum $k_{ini}$ and invariant mass $M_{ini}$
of the $Q\overline{Q}$ pair upon production, and the right side displays
the corresponding spectra with respect to 
$r_{fin}$, $k_{fin}$ and  $M_{fin}$ of the final $Q\overline{Q}$ upon
coalescence to charmonium bound-states.

The plots are rather self-evident and in line with intuitive expectations:
The initial $Q$ and $\overline{Q}$ are highly localized with a mean spatial
extent of $\sim 10^-{2}$ $fm$, whereas the final $Q\overline{Q}$ pair
has a much larger typical size, being of the order of the size
of the charmonium states (c.f. Table 1).
On the other hand,
the initial and final spectra in the relative $Q\overline{Q}$
momentum show a reverse pattern, namely a significantly narrower
distibution of the final $Q\overline{Q}$ momentum upon coalescence, 
being the cumulate
effect of gluon radiation that decreases that decreases the originally
obtained relative momentum of the $Q\overline{Q}$ when it was produced initially.
Similarly, the mass spectrum of the initial $Q\overline{Q}$, being
related to the distribution of $Q^2$ values upon production,
spreads out to masses well above 10 GeV, although exponentially damped.
The mass distribution of the final $Q\overline{Q}$, on the other hand,
is concentrated in its mean around the 3 - 4 GeV region, where the charmonium states lie.
$\frac{}{}$\vspace{1.0cm}
\begin{figure}
\epsfxsize=480pt
\centerline{ \epsfbox{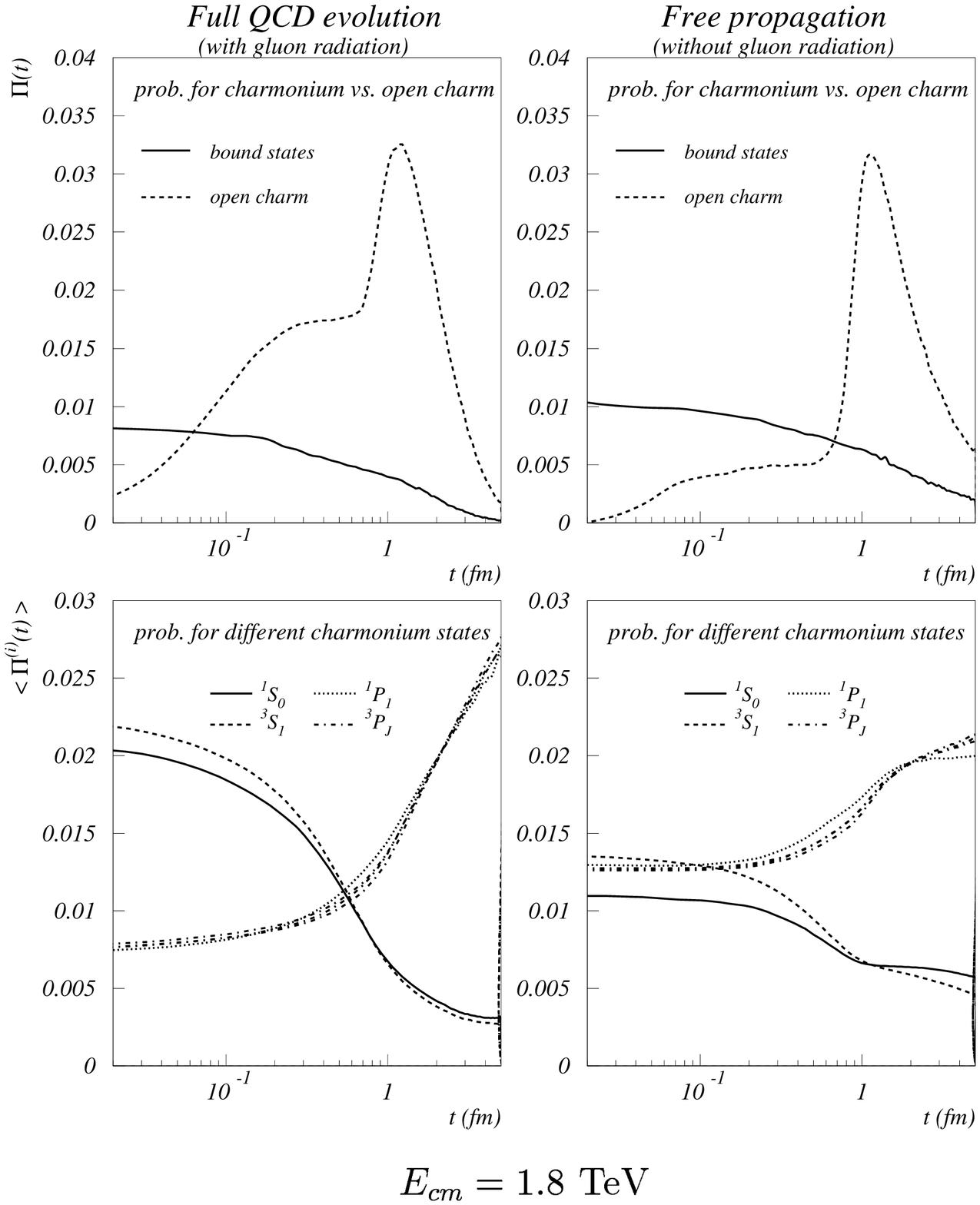} }
\vspace{-2.0cm}
\caption{
          Characteristics of the space-time evolution 
          of the `full QCD evolution' including gluon radiation,
          and the `free propagation' without gluon emission
          for  charm production at $p\bar{p}$ center-of-mass energy of 1.8 TeV.
          {\sl Top part}: time-development of  the probability for
          $Q\overline{Q}$ bound-state formation, 
          versus the probability for $Q$ and $\overline{Q}$
          to hadronize in open charm.
          {\sl Bottom part}: time-dependence of the individual coalescence 
           probabilities $\langle \Pi^{(i)}(t) \rangle$ for the spectrum of charmonium 
           states $i = ^1S_0, ^3S_1, ^1P_1, ^3P_J$.
           Included are only those $Q\overline{Q}$ pairs that indeed
           form bound-states, open charm production is subtracted.
         \label{fig:fig15}
         }
\end{figure}
$\frac{}{}$\vspace{1.0cm}
\begin{figure}
\epsfxsize=500pt
\centerline{ \epsfbox{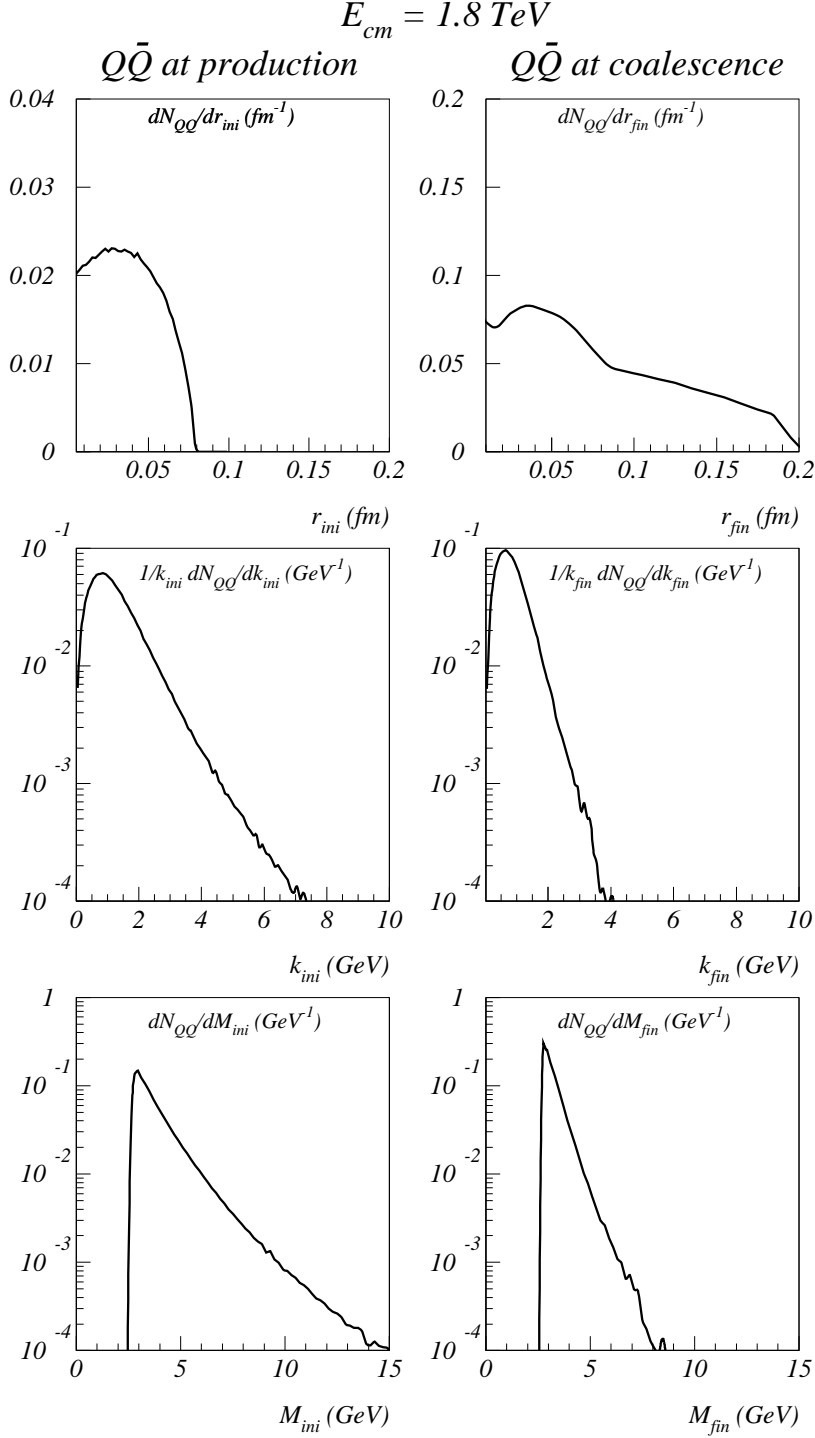} }
\vspace{-1.0cm}
\caption{
          Characteristics of the space-time evolution reflected in 
          initial and final $Q\overline{Q}$ spectra, for the case 
          of charmonium  $p\bar{p}$ center-of-mass energy of 1.8 TeV.
         {\sl Left panel}: Distributions in relative separation
          $r_{ini}=|{\bf r}_Q - {\bf r}_{\overline{Q}} |_{t=t_0}$
          relative momentum 
          $k_{ini}=|{\bf k}_Q - {\bf k}_{\overline{Q}} |_{t=t_0}$
          and invariant mass
          $M_{ini}=\sqrt{ (k^\mu_Q + k^\mu_{\overline{Q}})^2} |_{t=t_0}$
          of the {\it initial} $Q\overline{Q}$ pair upon production at $t=t_0$.
         {\sl Right panel}: Distributions in the corresponding
          quantities $r_{fin}$,  $k_{fin}$, $M_{fin}$
          of the {\it final} $Q\overline{Q}$ pair upon coalescence.
         \label{fig:fig16}
         }
\end{figure}
\newpage


\section{CONCLUDING REMARKS AN OUTLOOK}
\label{sec:section4}
\bigskip

\subsection{Summary}
\medskip

The scope of this work concerned
the development and application of
a space-time description for
the production of heavy quarkonium as $Q\bar{Q}$ bound-states in
hadron-hadron collisions.
The theoretical framework was developed on the basis of
the common factorization assumption of short-distance
PQCD production and evolution of heavy $Q\overline{Q}$
pairs and the long-distance bound-state formation process leading
to final-state quarkonia.
\smallskip

The separation of perturbative and non-perturbative momentum scales
(and space-time scales) was exploited to formulate
the dynamics of quarkonium production and the
associated cross-section
as the product of three functions
corresponding to $Q\overline{Q}$ production, parton cascade evolution, and
coalescence, repectively.
The practical calulation involved  
three corresponding elements:
firstly, the `hard' production 
of the initial $Q\bar{Q}$, directly or via gluon fragmentation and 
including both color-singlet and -octet contributions, secondly,
the subsequent development of the 
$Q\bar{Q}$ system   within a space-time generalization of
 the DGLAP parton-evolution formalism in position- and momentum-space,
and thirdly,
the actual formation of the bound-states being described by
the space-time dependent overlap between the $Q\bar{Q}$ pair 
and the spectrum of quarkonium wave-functions. 
\smallskip

This approach  was then  applied to
$pp$ ($p\bar{p}$) collisions in the energy range $\sqrt{s} = $ 30 GeV - 14 TeV,
corresponding to the range of center-of-mass energies between 
ISR, Tevatron, and LHC experiments,
by simulating the space-time development of the complete particle system
of a collision event, using a suitably extended version of the computer program VNI.
Comparison with experimental data from ISR and Tevatron for both charmonium and bottonium
production revealed very good agreement of the model calculations,
essentially without the necessity of introducing tunable, free parameters.
Details of the underlying dynamics that lead to the decent
{\it post}dictions of ISR and Tevatron data (and {\it pre}dictions for LHC),
were analyzed in particular with respect
to the relevance of gluon radiation, the contributions of
color-singlet and color-octet $Q\overline{Q}$ systems, and the characteristics
of the space-time structure.
\bigskip

\subsection{Future perspectives}
\medskip

Finally some remarks on future perspectives.
The good agreement with available experimental data of 
the present approach to quarkonium production in $pp$ or $p\bar{p}$
collisions, as well as its transparent physics concept, suggests 
to extent applications and tests of the model to other
collision systems.

\begin{itemize}
\item
{\it pion-proton collisions} are experimentally well
studied too in terms of quarkonium production (although  at lower
energies), observed with even 
distinctly larger yields, partly  due to the different valence quark content
of the initial state.
It would be interesting to test whether the model advertized here is capable
of describing these reactions equally well.
\smallskip

\item
{\it proton-nucleus collisions} are another extension for which again quite some
experimental data exist for various energies and different nuclei.
The novel feature here, as compared to hadron-hadron collisions, is
the propagation of the $Q\overline{Q}$ pair in nuclear matter,
that is, before forming a bound-state, $Q$ and/or $\overline{Q}$ 
may experience scatterings
off partons belonging to close-by nucleons in the traversed nucleus. 
As a consequence, one expects the quarkonium yield 
to be reduced due to the deflection
of $Q$ and/or $\overline{Q}$ and the resulting failure of coalescence.
\smallskip

\item
{\it nucleus-nucleus collisions}, finally, are a most interesting ground for
studying quarkonium production in QCD matter, in
particular since  it has become appreciated as one
of the best experimental probes  for indirect detection 
of a quark-gluon plasma, a high-density and
high-temperature state of deconfined partonic matter
for which first hopeful signals have been extracted recently 
in lead-on-lead collisions at the CERN SPS facility.
Compared to proton-nucleus collisions, in heavy-ion reactions
several new physics issues arise. For example,
the large size of the nuclear collision system makes it likely
that the traversed distance of a heavy $Q\overline{Q}$ system
can extend  over   several $fm$ so that not only
the $Q$ and $\overline{Q}$ are likely to re-interact before
they may coalesce, but also the coalesced $Q\overline{Q}$ bound-states
can re-collide and being broken up again (so-called `dissociation').
Moreover, 
in central collisions of heavy nuclei at sufficiently large energy per
nucleon,
it is expected \cite{msrep} that
copious parton materialization and mini-jet production through
very frequent parton collisions, plus further parton multiplication
through gluon emission induced by partonic re-scatterings,
may create very quickly (with in a fraction of a $fm$) a very dense
and highly excited partonic environment. Whether or not
this QCD matter approaches an equilibrium state, i.e. some sort
of quark-gluon plasma, the properties of the hot and dense
parton matter are of their own fundamental interest.
In this context, quarkonium production is certainly a pre-destined probe
to study these properties, as mentioned before.
Both the yield of formed $Q\overline{Q}$ bound-states and
the survival probability of resulting quarkonium states are likely
to be suppressed, the degree of suppression being an indicator
of the density and temperature of the nuclear matter.
It is clear that
here  the space-time picture is an inevitable
necessity to  study quarkonium production in high-density QCD, where 
accounting for the dynamically
evolving space-time structure of the collisions is most important, and where
the time-dependent spatial profile of the particle density is the
key quantity that determines the fate of quarkonium.
Since the local density is created by
the internetted space-time structure of many parton cascades
evolving close-by in the central collision region, it
the only way of `calculating' the dynamics is by rigorous
accounting for both
the space-time variables and
the energy-momentum variables for each particle present at a given time during
the collision.
Here lies a real challenge for the model advocated in this paper.
The technicalities of extending the present status to nuclear collisions
are  straightforward, though non-trivial, and work is in progress.
\end{itemize}
\bigskip
\bigskip
\bigskip
\bigskip

\section*{ACKNOWLEDGEMENTS}
This work was to large extent inspired and driven by numerous discussions
with Sidney H. Kahana and David E. Kahana, who also were involved in
a revision of an initial draft of the manuscript.
Thank you very much.
Furthermore, I would like to thank Eric Braaten for guidelining suggestions,
and Michelangelo Mangano for encouragement as well as supplying
the  Tevatron data in ready-to-plot format.
This work was supported in part by the D.O.E under contract no.
DE-AC02-76H00016.

\end{document}